\documentclass[12pt]{article}
\usepackage{pic04}
\usepackage{hyperref}
\usepackage{url}
\usepackage{graphicx}

\begin{document}

\title{\bf SEARCH FOR NEW PHENOMENA AT COLLIDERS}
\author{Elem\'er Nagy \\  
       for the CDF, D\O, H1 and ZEUS Collaborations \\
{\em CPPM, CNRS-IN2P3, and Universit\'e de la M\'editerran\'ee, Marseille, France}}
\maketitle

%
%
%
%
%
%
%

\baselineskip=14.5pt
\begin{abstract}
Recent results on searches for new phenomena on the Tevatron and
HERA colliders are reviewed.
\end{abstract}

\newpage

\baselineskip=17pt

\section{Introduction}

In spite of the great succes of the Standard Model (SM) we still
do not have a unified description of all the
four known forces in a finite, renormalizable theory.
Therefore new phenomena are expected beyond the SM at some
energy scale $M_X$. 
Today only elements of this ultimate theory are available and are
proposed for experimental
tests. In the present paper I have chosen an arbitrary list of these tests
addressed by two active colliders, Tevatron and HERA:
{\it Extra Dimensions (ED), Supersymmetry (SUSY), Z', Leptoquarks,
Beyond SM Higgs bosons, and Substructures of quarks and leptons.}
The confrontation of these models with experiment
is  complemented by a {\it systematic search} for departures from the SM.
Concerning the scale $M_X$, its natural value would be the
Planck-mass, $M_{Pl}$ if one is aiming at unification with gravity. 
However the very large value of $M_{Pl}$, in comparison with the electro-weak scale 
leads to ``unnatural'' fine tuning
of scalar masses, known as the {\it hierarchy problem}. Some of the above
topics present solution to this problem.

Frequently, the same event topology, like e.g. high mass lepton pairs,
allows to test several theoretical models.

In the Tevatron collider at Fermilab, near Chicago, protons collide with antiprotons
at a center-of-mass energy of 1960 GeV. Two experiments are collecting data,
CDF and D\O . Both experiments as well as the Tevatron underwent substantial
upgrades in the last decade. In the new data taking periode (Run II) started 
in spring 2001, the Tevatron has delivered
already 3 times more luminosity as in Run I. The detector upgrades  allow
efficient operation at the increased luminosity, provide extended coverage
of subdetectors and in the case of D\O\ a completely new central tracking. 
In this paper only the
most recent results from Run II will be reviewed making use of typically 200 pb$^{-1}$
integrated luminosity.

In the HERA collider at DESY, near Hamburg, positrons or electrons collide with protons
at 300 (or 318) GeV energy in the center of mass. The data are collected by
two experiments, H1 and ZEUS. Here again, the collider as well as the two
detectors have been upgraded. The new periode of data taking (HERA II) has started
recently with the goal to collect 10 times more luminosity as in HERA I
and with a possibility of using longitudinally polarized electron beam.
In this paper only results from HERA I will be covered since results 
with comparable statistics from the new data taking periode 
hasn't been published yet.

All limits are quoted at 95\% confidence level. References for theoretical
models can be found in those quoted for experimental results.

\section{Extra Dimensions}

As Theodor Klauza has shown the first time, almost a century ago, extra dimensions (ED)
can provide a framework for unification with gravity. ED's were supposed to be compact
of a very small size (Oscar Klein) since we don't sense them in our everyday experiences. Particles
propagating in compact ED's have higher mass replica's, the so-called Klauza-Klein (KK) states.
Recently, ED's revived interest since their existence is needed in string theories,
and also because it was pointed out that their size can be large (LED), and therefore
accessible for experimental verification. Moreover, LED's explain that gravity is 
only apparently weak and the scale of new physics can be much lower than the Plank mass,
thereby avoiding the hierarchy problem.

ED's can be tested experimentally on colliders either by virtual effects or direct
emission of the KK-states. 

\subsection{Search for LED's}

\begin{figure}[htbp]
  \centerline{\hbox{ \hspace{0.2cm}
    \includegraphics[width=6.5cm]{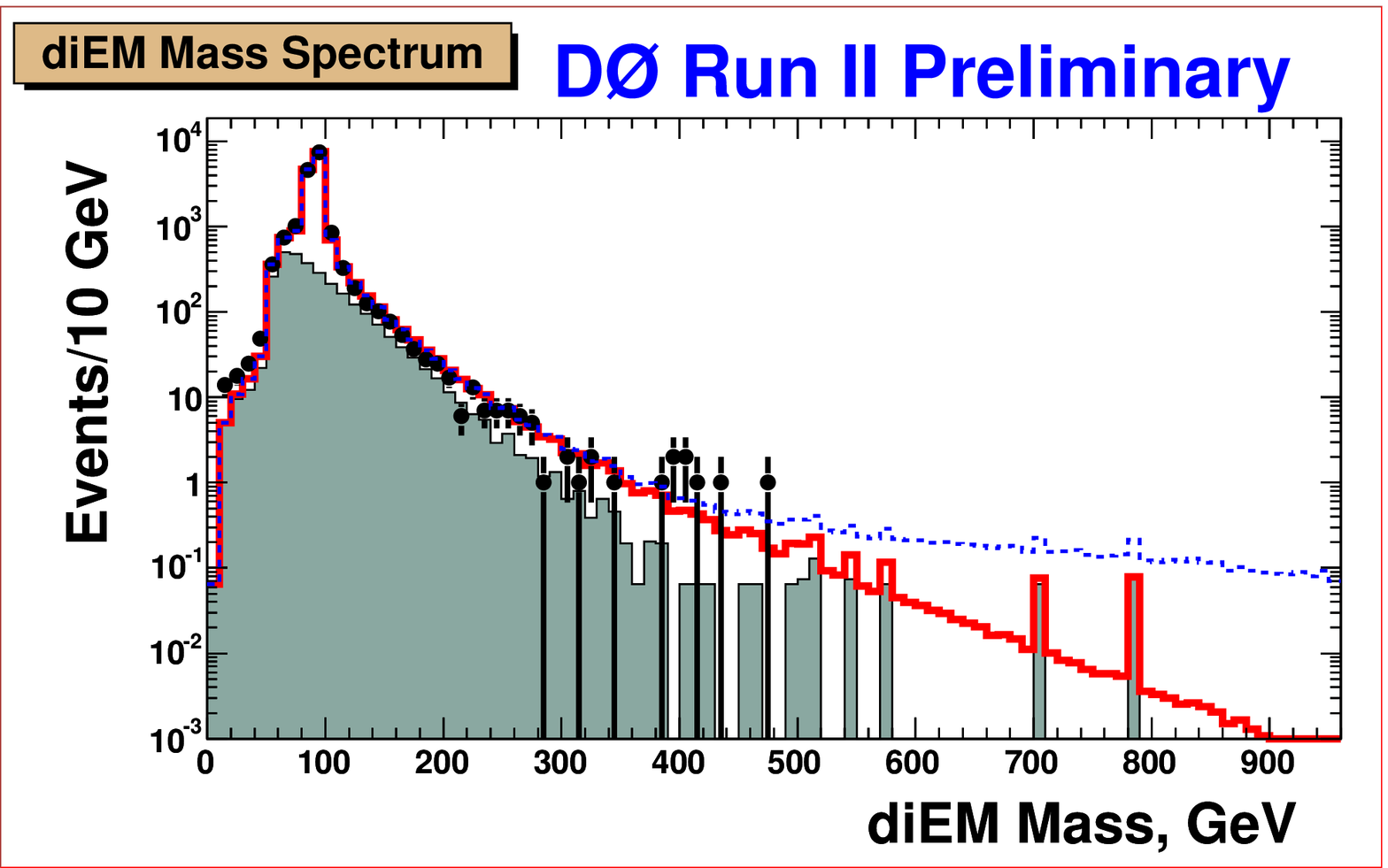}
    \hspace{0.3cm}
    \includegraphics[width=6.5cm]{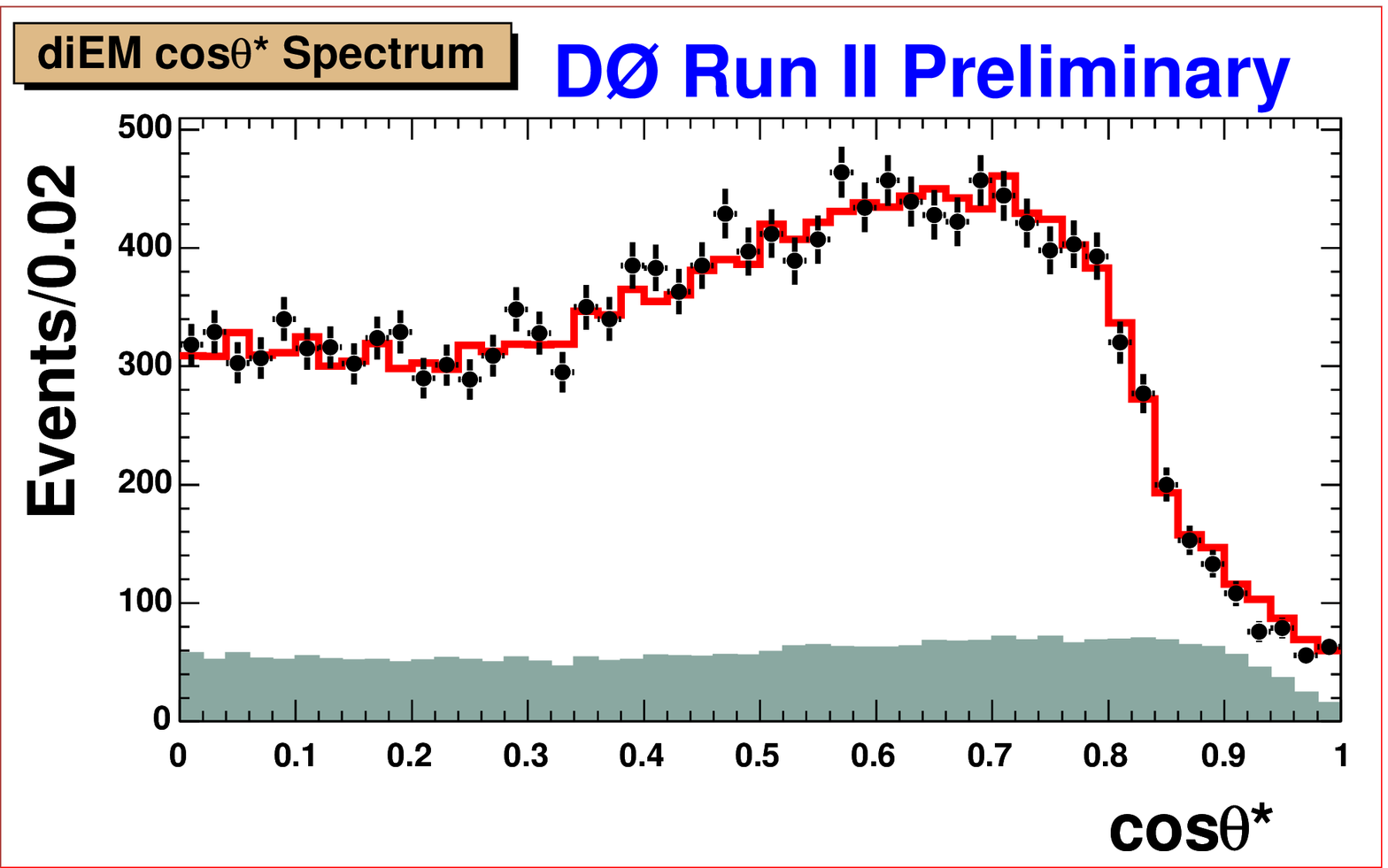}
          }
         }
 \caption{\it
      The invariant mass (left) and  angular distribution (right)
of di-electron and di-photons obtained by D\O\ . Points with error
bars are the data; light filled histogram represents the
instrumental background, solid line shows the fit to the sum
of the instrumental background and SM predictions. The dashed line shows
the effect of the LED signal for $\eta_G=0.6$.
    \label{fg:LEDdiemD0} }
\end{figure}

\begin{figure}[htbp]
  \centerline{\hbox{ \hspace{0.2cm}
\includegraphics[width=7.0cm]{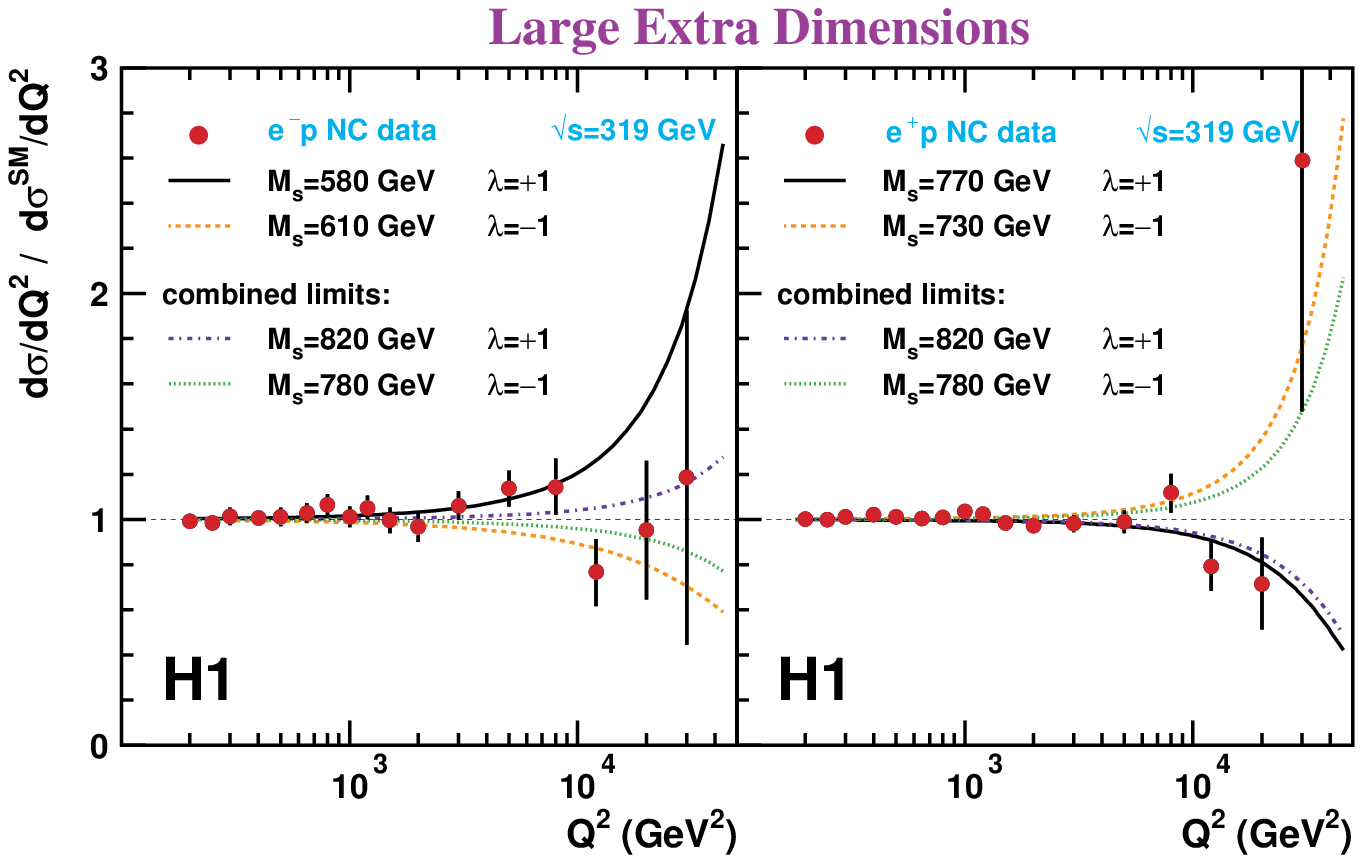}
    \hspace{0.3cm}
    \includegraphics[width=6.0cm]{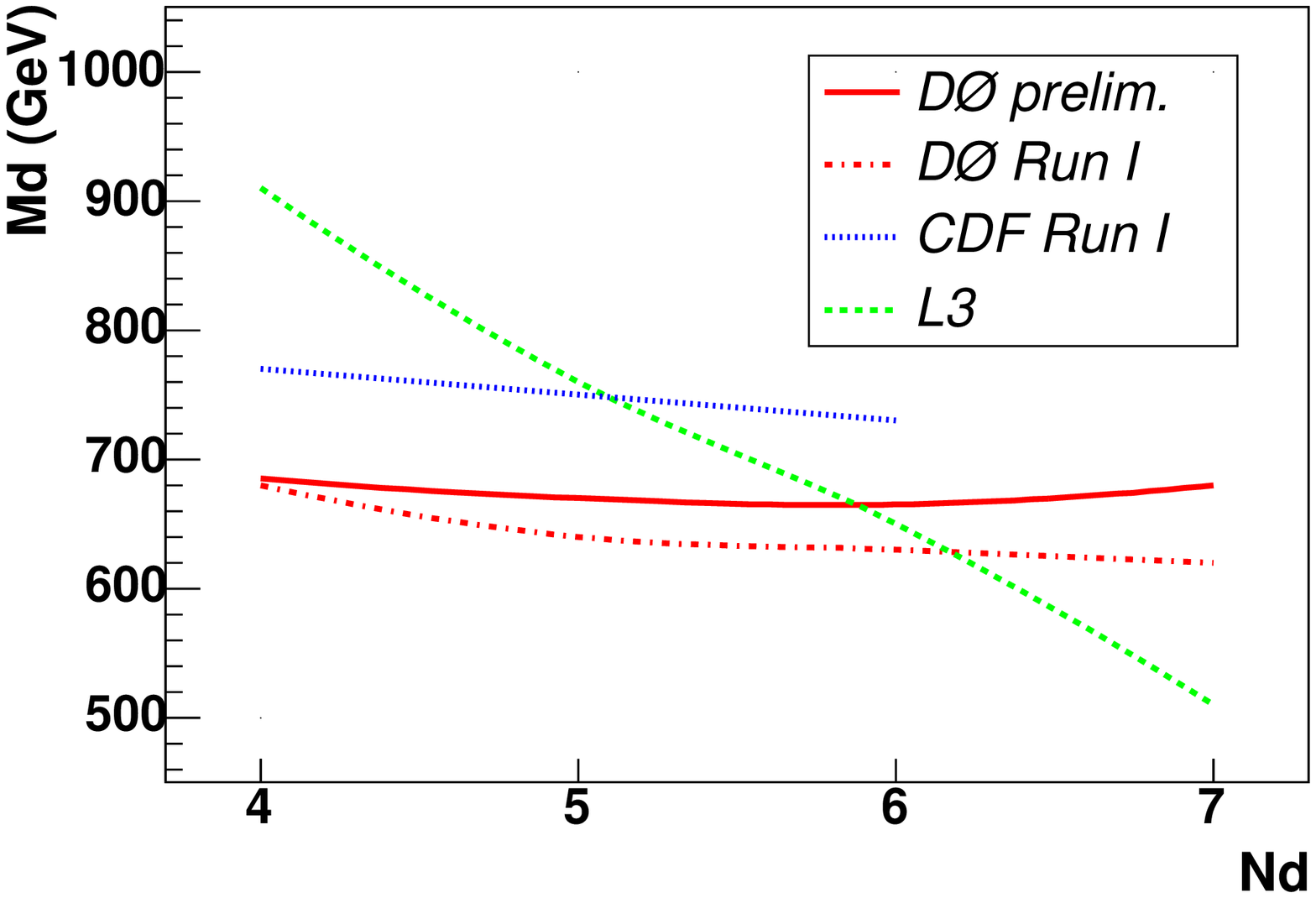}
        }
        }
 \caption{\it
    (a) NC cross sections obtained by H1 at $\sqrt{s}$ = 319 GeV normalized to the SM
expectation (left).   
    (b) The present limits on the fundamental scale
of the gravitation ($M_d$) as a function of the number of ED's (right)
obtained at the Tevatron in the monojet channel.   
    \label{fg:LEDH1} }
\end{figure}

LED's can be of macroscopic size if only gravity is supposed to propagate
in them. Indeed, gravity experiments have tested Newton's law only
down to $\sim 10^{-4}$ m distances. 
The {\it virtual effects of the graviton KK-states} show up at hadron colliders e.g. as deviation in the
distribution of the invariant mass, $M$ and that of the angle $\theta^*$ in the center of mass
of di-lepton and di-photon states. D\O\  has studied
these distributions not separating the di-electron and di-photon states~\cite{D0Results}$(i)$.
The expected deviation can be parametrized as:    
\begin{equation}
\frac{d^2\sigma}{dMd\cos\theta^*} = f_{SM}+f_{int}\eta_G+f_{KK}\eta_G^2, 
\label{eq:LEDdiem}
\end{equation}
where $f_{SM}$, $f_{KK}$ and $f_{int}$ are the contributions due to the SM, KK gravitons and
the interference between them. The constant $\eta_G$ contains the fundamental scale of the
gravitation, $M_S$, given e.g. by Hewett in the following form:
\begin{equation}
\eta_G = \frac{2\lambda}{\pi}\frac{1}{M_S^4};\ \ \lambda=\pm1.
\label{eq:Hewitt}
\end{equation}
D\O 's result is displayed in Fig.~\ref{fg:LEDdiemD0}. The data points follow the expected
background. The absence of events at high mass allowed to obtain lower
limits for $M_S=1.22\ (\lambda=+1)$ and $M_S=1.10\ (\lambda=-1)$.
Similar conclusions have been obtained by CDF~\cite{CDFResults}: 
$M_S=0.961\ (\lambda=+1)$ and $M_S=0.987\ (\lambda=-1)$.
At HERA one has compared the inclusive neutral current (NC)
deep inleastic cross sections to 
that expected from the SM. As no deviation has been found (see e.g.
Fig.~\ref{fg:LEDH1}$a$) both experiment established limits on $M_S$
(using $\eta_G = \frac{\lambda}{M_S^4}$), 
H1: $M_S=0.82\ (\lambda=+1)$ and $M_S=0.78\ (\lambda=-1)$~\cite{H1Results}$(i)$,  
ZEUS: $M_S=0.78\ (\lambda=+1)$ and $M_S=0.79\ (\lambda=-1)$~\cite{ZEUSResults}$(i)$. 

{\it Direct emission of KK gravitons} have been searched for on the Tevatron
by both collaborations. The signature is a monojet, i.e. a jet
of high transverse energy ($E_T$) accompanied by a large missing transverse
energy (MET). No significant number of such events above the expected
backgorund has been found. The present limits on the fundamental scale
of the gravitation ($M_d$) as a function of the number of ED's is
given in Fig.~\ref{fg:LEDH1}$b$~\cite{D0Results}$(ii)$. 

\subsection{Search for TeV$^{-1}$ size LED}

D\O\ has searched for TeV$^{-1}$ size LED in a model where fermions are confined in
the ordinary 3$d$ world, in contrast to gauge bosons which can propagate in
an extra compact dimension. Such a possibility can lead to spectacular minima
and secondary maxima in the di-lepton invariant mass distribution due to the
interference of the KK states of gauge bosons. The absence of such behaviour
in the di-electron channel allowed to set a lower limit of 1.12 TeV for the
inverse size of the ED~\cite{D0Results}$(iii)$. 
   
\subsection{Search for Randall-Sundrum resonances}

\begin{figure}[htbp]
  \centerline{\hbox{ \hspace{0.2cm}
\includegraphics[width=6.5cm]{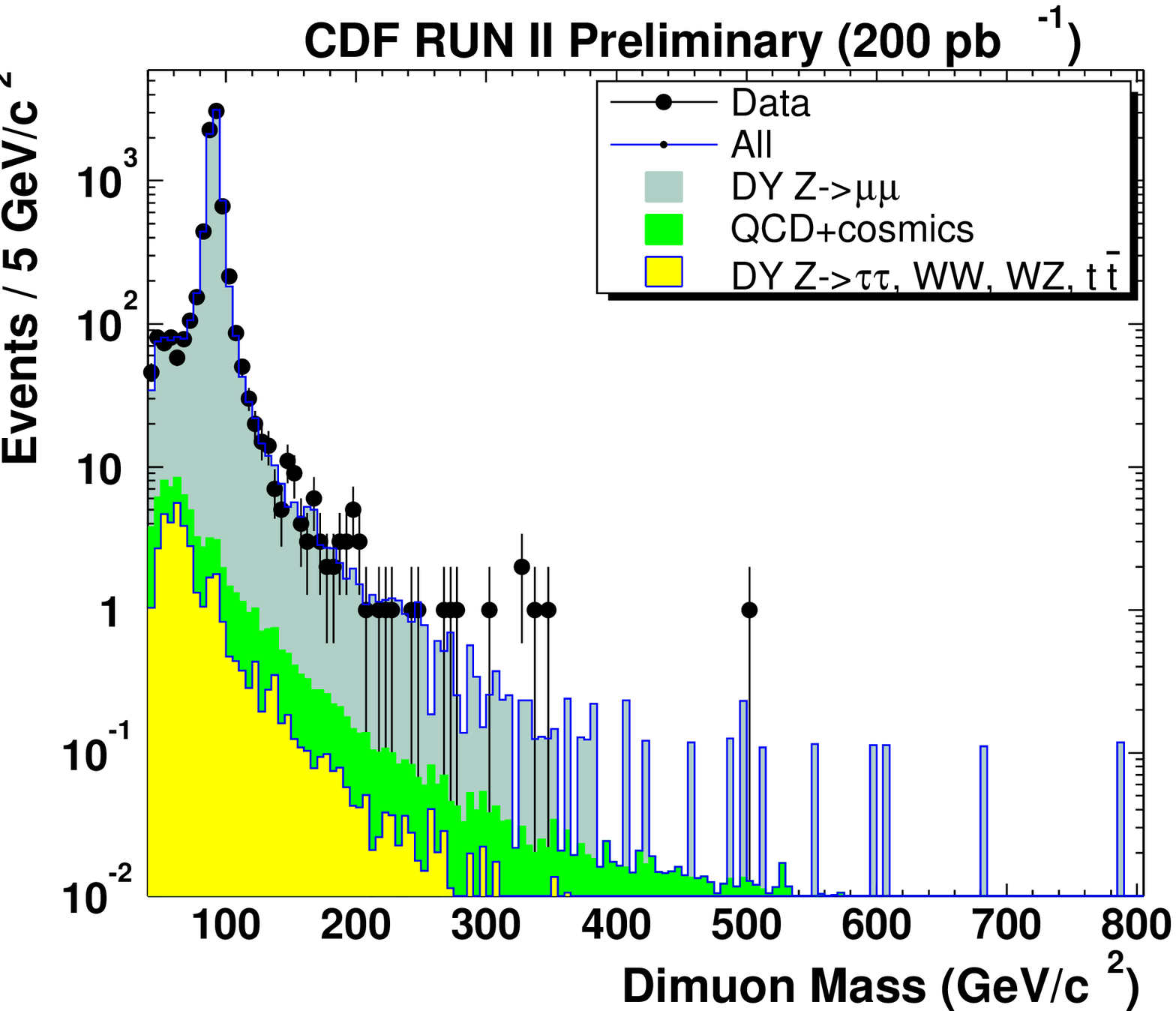}
    \hspace{0.3cm}
    \includegraphics[width=6.5cm]{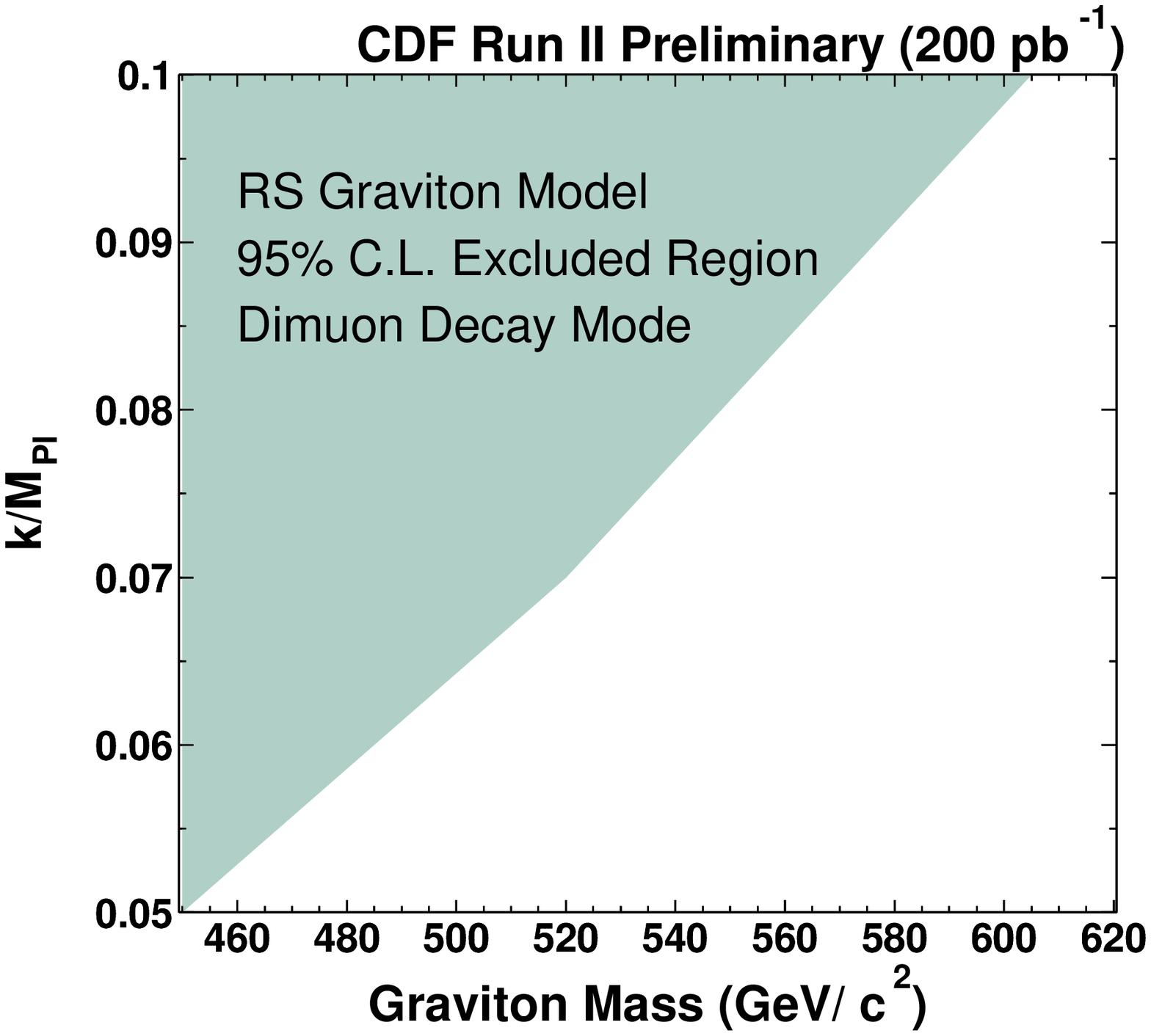}
        }
        }
 \caption{\it
(a) Invariant mass distribution of opposite sign muon pairs observed
by CDF (left) together with the estimated background. 
(b) Limit on the mass of Randall-Sundrum graviton resonances
as a function of the model parameter $k$ (right).    
    \label{fg:CDFdimuMass} }
\end{figure}

CDF has searched for graviton resonances predicted by the Randall-Sundrum model
in the mass distribution of di-lepton pairs. Fig~\ref{fg:CDFdimuMass}$a$ shows
the data and the expected background for the muon pairs. Since data and
background are in agreement and no resonant structure is observed, CDF has
established limits on the resonance mass as a function of the model parameter,
$k$ (Fig.~\ref{fg:CDFdimuMass}$b$)~\cite{CDFResults}.

\section{Supersymmetry (SUSY)}

SUSY is a symmetry of Nature with respect to interchange of
bosons and fermions. It is a basic ingredient for unification
with gravity (e.g. in Superstring/M-theories). It is also the
only non-trivial extension of the Lorentz-Poincar\'e group.
Moreover, it presents an elegant solution for the hierarchy
problem. In the minimal supersymmetric extension of the SM (MSSM),
every SM particle has a SUSY partner whose spin differs by
$\pm 1/2$. $R$-parity, defined as $R=(-1)^{3B+2L+S}$, where
$B$, $L$ and $S$ are the baryon number, lepton number and the
spin, is $+1$ for SM particles and is $-1$ for their SUSY
partners. Apart of the SUSY partners a second Higgs doublet
is also needed in the MSSM. Its mass parameter represents the
only additional parameter in the theory if SUSY were an exact
symmetry.

SUSY is however a broken symmetry, at least at the electro-weak
energy scale, which introduces more than a hundred new parameters.
With additional hypotheses one can reduce this number to a
managable size: 5 ($m_0$, $m_{1/2}$, $\tan\beta$, sgn$\mu$
and $A_0$) for the gravity mediated symmetry breaking
model, mSUGRA, and 6 ($\Lambda$, $M_m$, $N_5$, $\tan\beta$, sgn$\mu$,
and $C_{grav}$) for the gauge mediated symmetry breaking
model, GMSB, considered here~\cite{SUSY}.

$R$ parity is approximately conserved in order to avoid $B$ and $L$
violating processes. In this case SUSY partners are pair-produced
and the lightest SUSY particle (LSP) is stable. Since it is believed
to be neutral and it interacts weakly, the basic experimental signature
for SUSY is large MET. It is accompanied by several leptons and jets
from cascade decays of SUSY particles. The main SM background
is $t\bar t$ and gauge boson pair production.

However small  $R$ parity violation cannot be excluded which allows
single resonant formation of the SUSY partners and gives rise to more
leptons and jets in the final state due to the decay of the LSP.
At Tevatron both $R$ parity conserving (RPC) and $R$ parity violating (RPV)
processes can be studied. HERA is competitve only in RPV SUSY searches.  

\subsection{RPC SUSY searches at the Tevatron}

For this kind of searches the golden channel is chargino-neutralino
pair production where the large MET is accompanied with several leptons.
D\O\ has used the $e-\mu-l$, $e-e-l$ and $\mu^{\pm}-\mu^{\pm}$ signatures,
where $l$ stands for a lepton having a charged isolated track. LEP has already
set stringent limits in the mSUGRA parameter space for these processes.
D\O\ has therefore investigated the region of $72 \le m_0 \le 88$ GeV,
$165 \le m_{1/2} \le 185$ GeV, $\tan\beta = 3$, $\mu>0$, $A_0=0$. 
This region is situated above the LEP limits and is characterized by the
mass relations
$m_{\chi_1^{\pm}}\approx m_{\chi_2^0}\approx 2m_{\chi_1^0} \approx m_{\tilde l}$
offering the highest discovery potential. In all three channels
the number of candidate events
has been found compatible with that estimated from the SM background
in the kinematical regions where the expected signal is dominant. This
allowed to obtain upper limit of appr. 0.5 pb for the production cross
section times branching ratio above the chargino mass limit established
by LEP ($\approx$ 103 GeV). This limit is a considerable improvement with
respect to that obtained in Run I but is slightly higher than the mSUGRA
prediction ($\approx$ 0.3 pb)~\cite{D0Results}$(iv)$.

\begin{figure}[htbp]
  \centerline{\hbox{ \hspace{0.2cm}
\includegraphics[width=4.5cm]{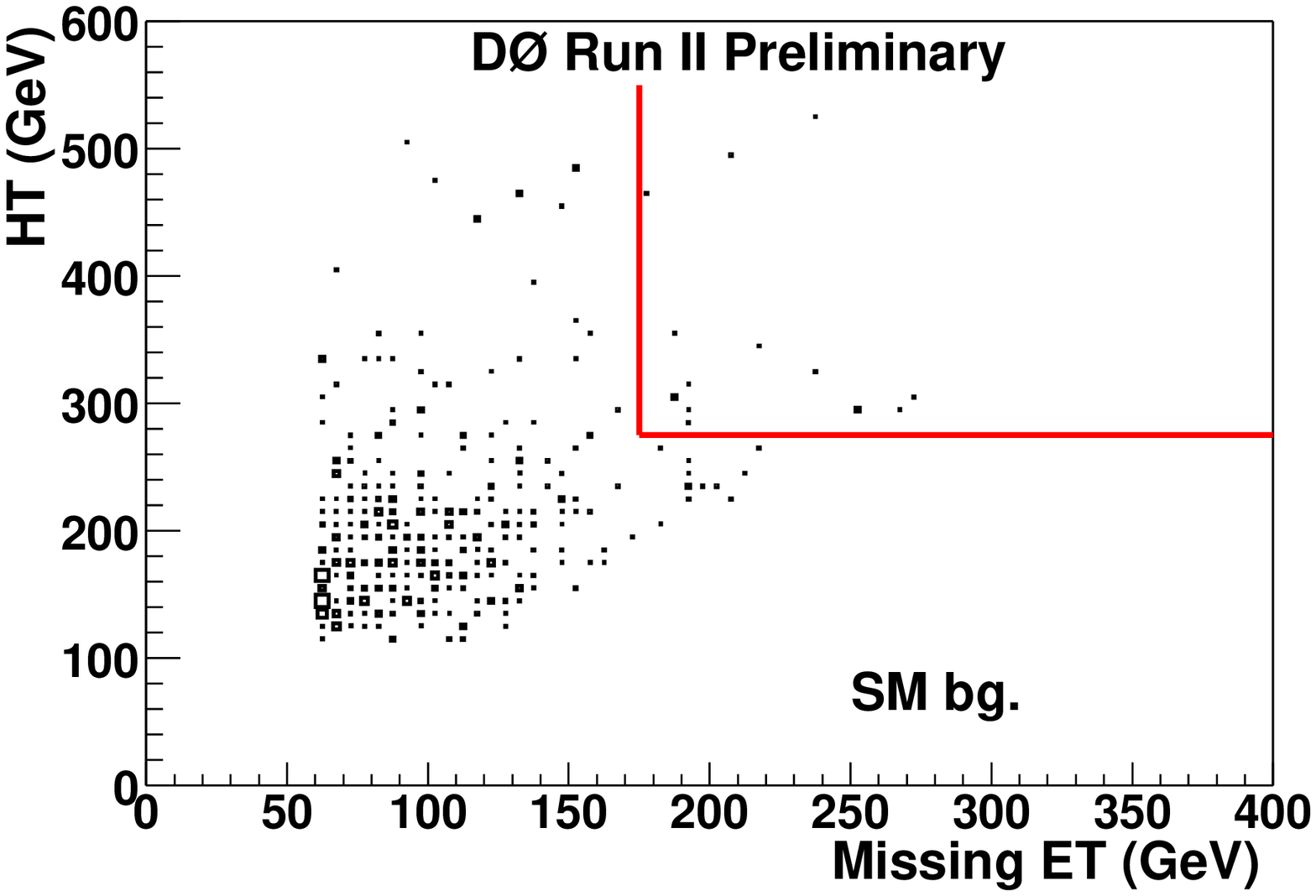}
    \hspace{0.3cm}
    \includegraphics[width=4.5cm]{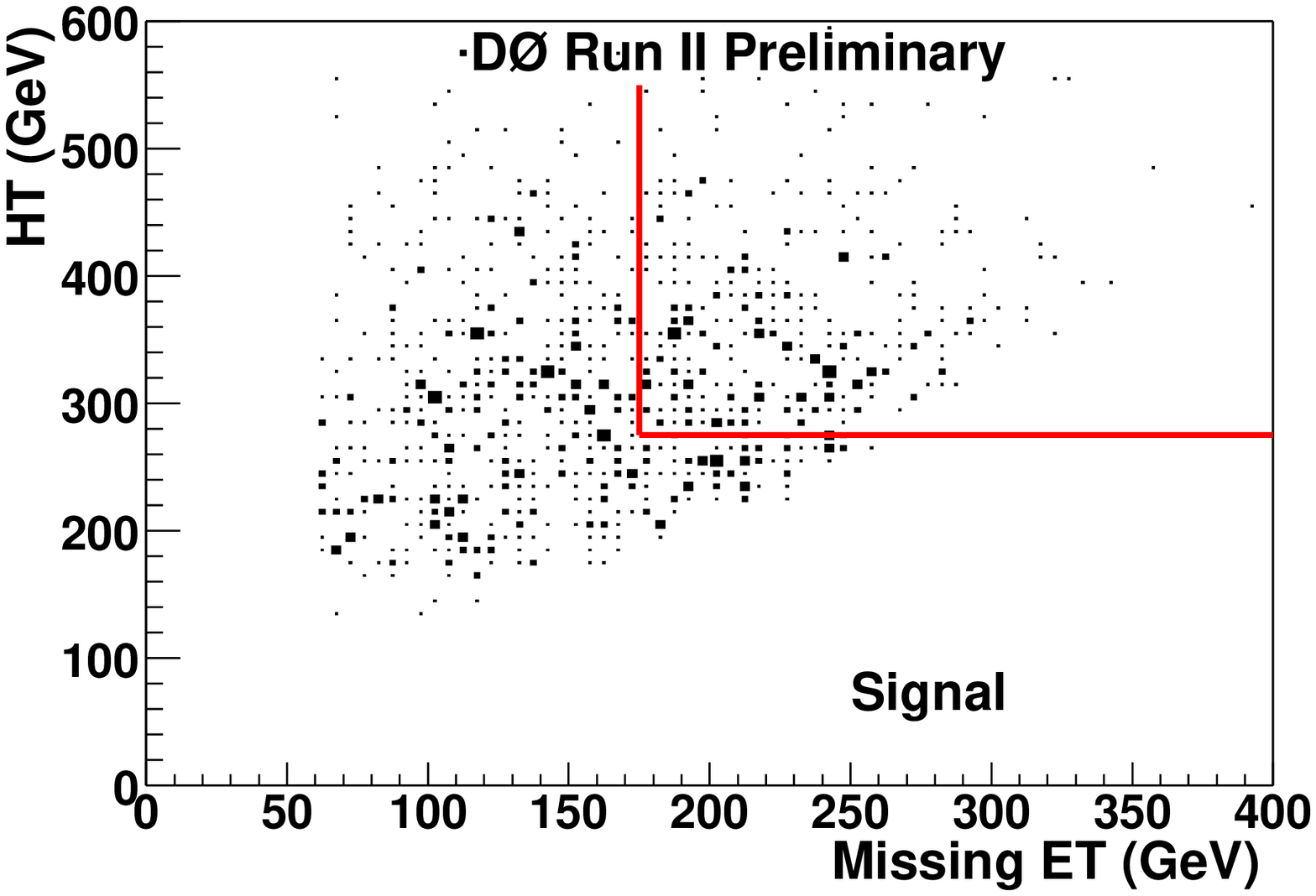}
    \hspace{0.3cm}
    \includegraphics[width=4.5cm]{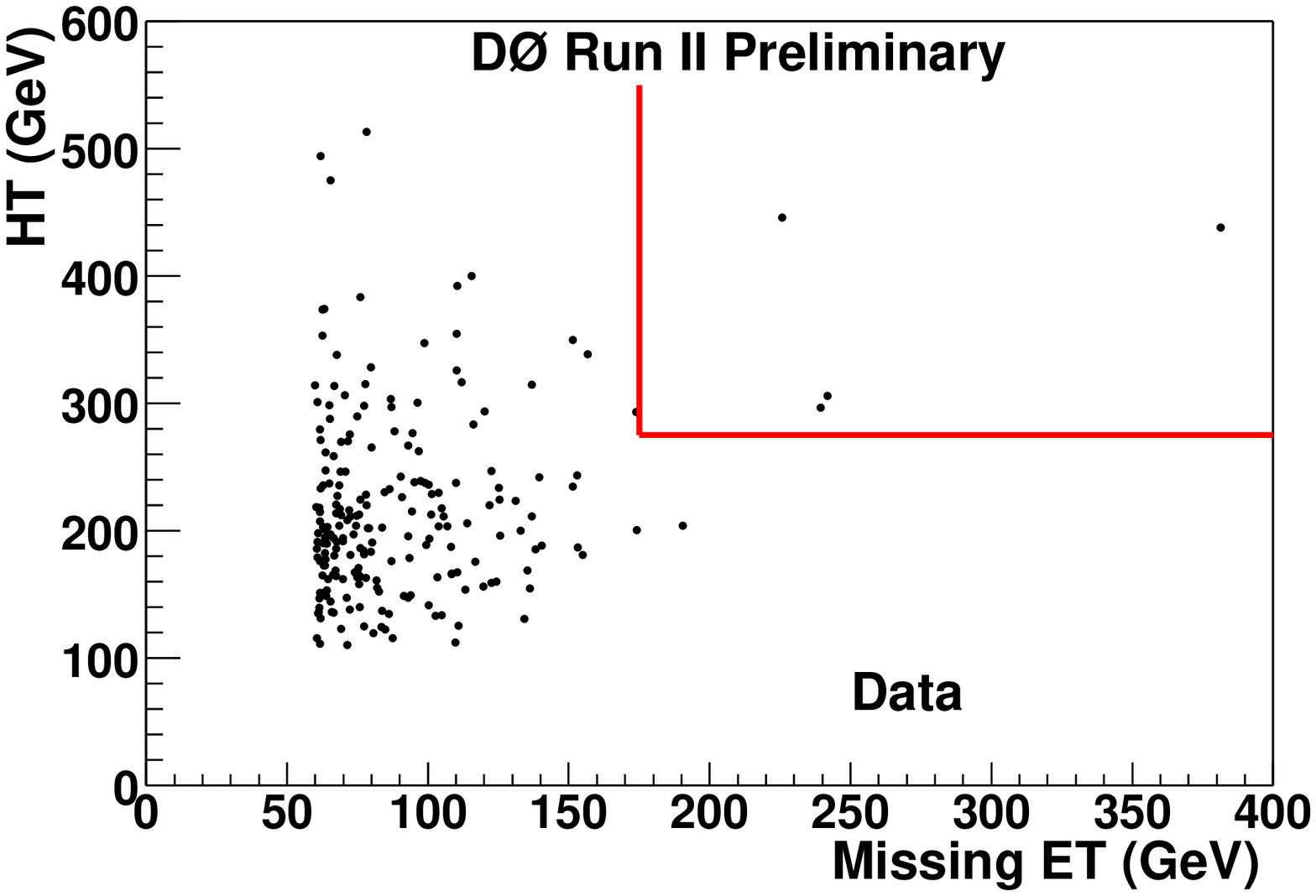}
        }
        }
 \caption{\it
Expected SM background (left), $s$quark/gluino signal (middle) and observed
data by D\O\ in the HT vs MET plot (right).    
    \label{fg:mSUGsqglSMbg} }
\end{figure}

The signature of $s$quarks and gluinos are multiple jets and MET. As shown 
in Fig~\ref{fg:mSUGsqglSMbg} signal and background can be well separated
using MET and HT, where this latter is the scalar sum of the jets.   
Since the data points are dominantly in the background region, $s$quarks and
gluinos have been excluded by D\O\ below masses $m_{\tilde q}<292$ and
$m_{\tilde g}<333$ GeV, respectively, in the mSUGRA parameter space of
$m_0 = 25$ GeV,
$100 \le m_{1/2} \le 140$ GeV, $\tan\beta = 3$, $\mu<0$, $A_0=0$~\cite{D0Results}$(v)$.

\begin{figure}[htbp]
  \centerline{\hbox{ \hspace{0.2cm}
\includegraphics[width=5.5cm]{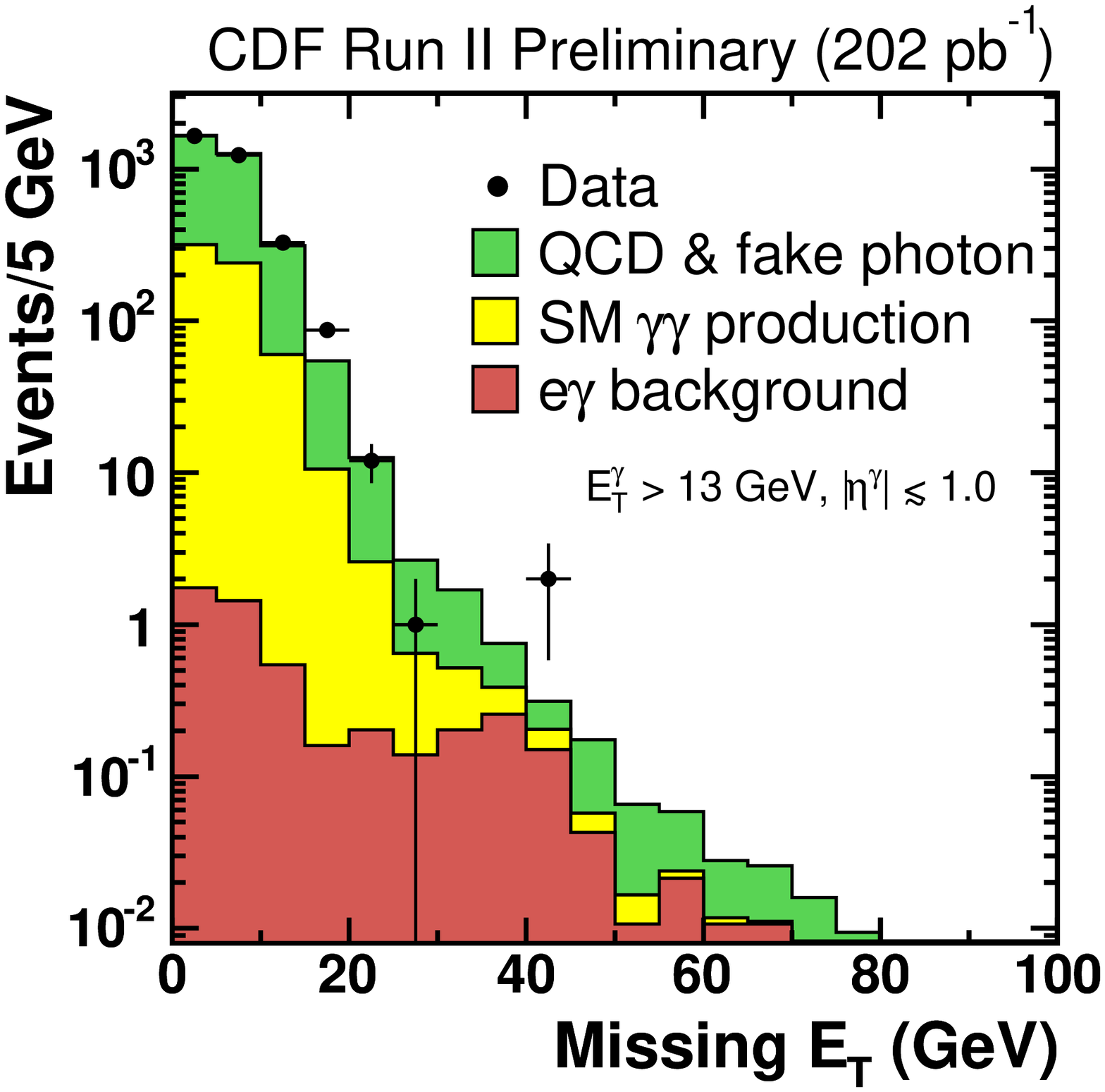}
    \hspace{0.3cm}
    \includegraphics[width=7.5cm]{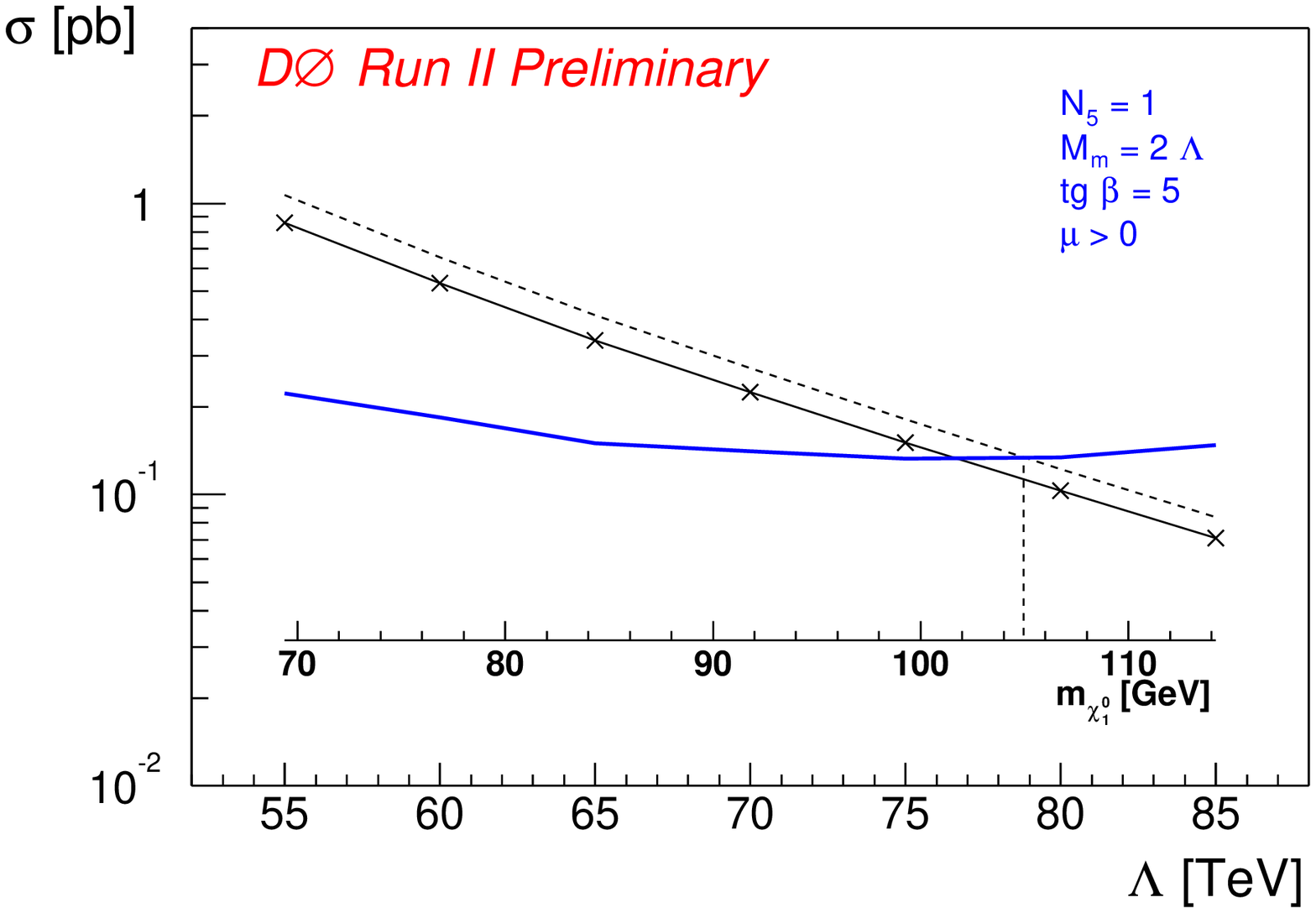}
        }
        }
 \caption{\it
(a) MET distribution of events accompanied by 2 isolated photons observed by CDF
(left). (b) Cross section limit of inclusive 2 photon events with MET $>$ 40 GeV,
obtained by D\O\ and the theoretical cross section of the 
GMSB model (dotted line indicates NLO
calculation).      
    \label{fg:CDF2gamMET} }
\end{figure}

In the GMSB model the LSP is a very light gravitino.
Assuming that the next LSP is the lightest neutralino, which decays promptly
to a photon and a gravitino, CDF~\cite{CDFResults} and D\O ~\cite{D0Results}$(vi)$ 
searched  for events with 2 isolated
photons accompanied by MET. Both experiments found that the MET distribution
was in agreement with the expected background (Fig~\ref{fg:CDF2gamMET}$a$). 
This allowed to set limit on
the SUSY breaking scale $\Lambda > 78.8$ TeV as shown e.g. in Fig~\ref{fg:CDF2gamMET}$b$. 

\subsection{RPV SUSY searches}

\begin{figure}[htbp]
  \centerline{\hbox{ \hspace{0.2cm}
\includegraphics[width=6.5cm]{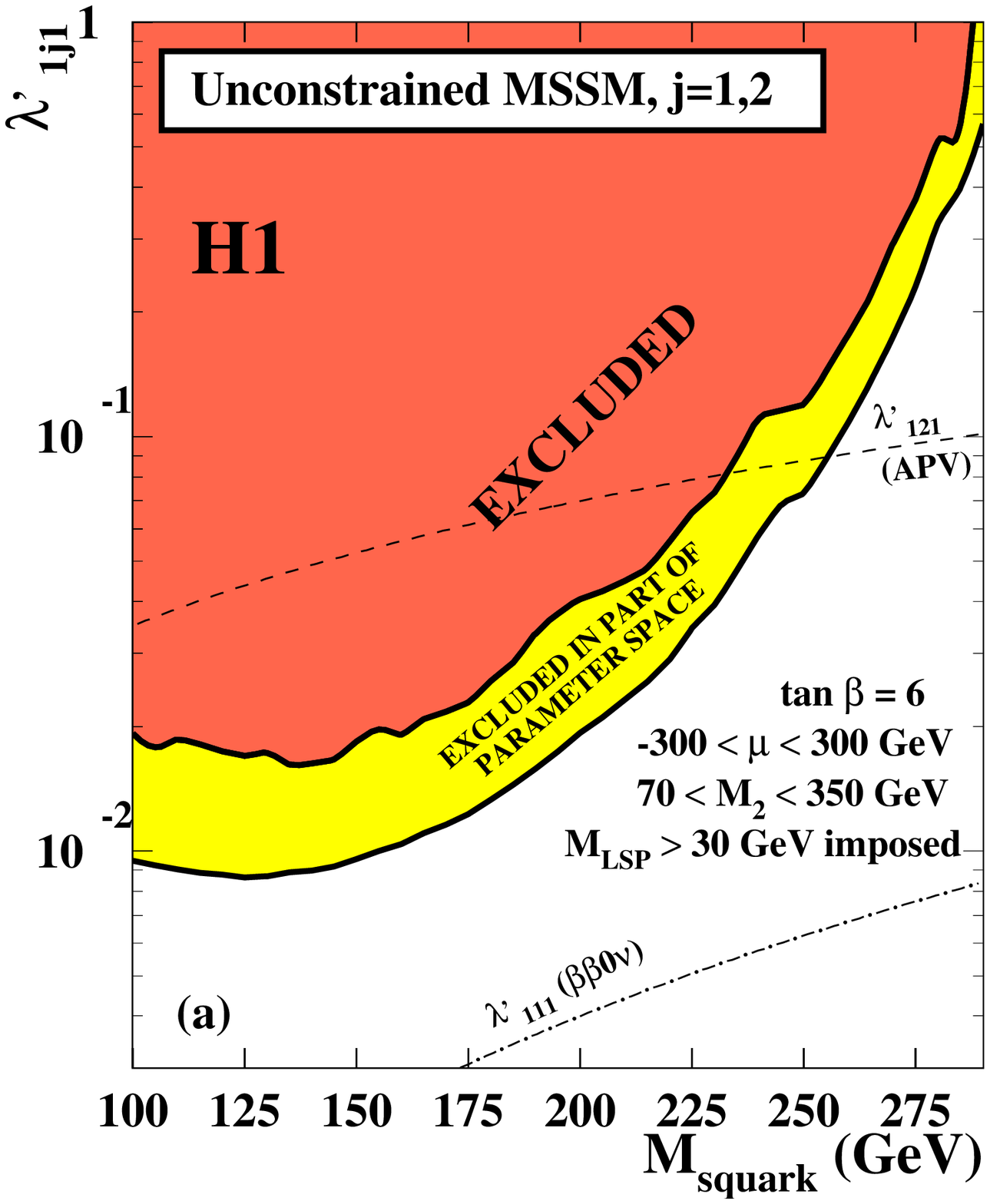}
    \hspace{0.3cm}
    \includegraphics[width=7.0cm]{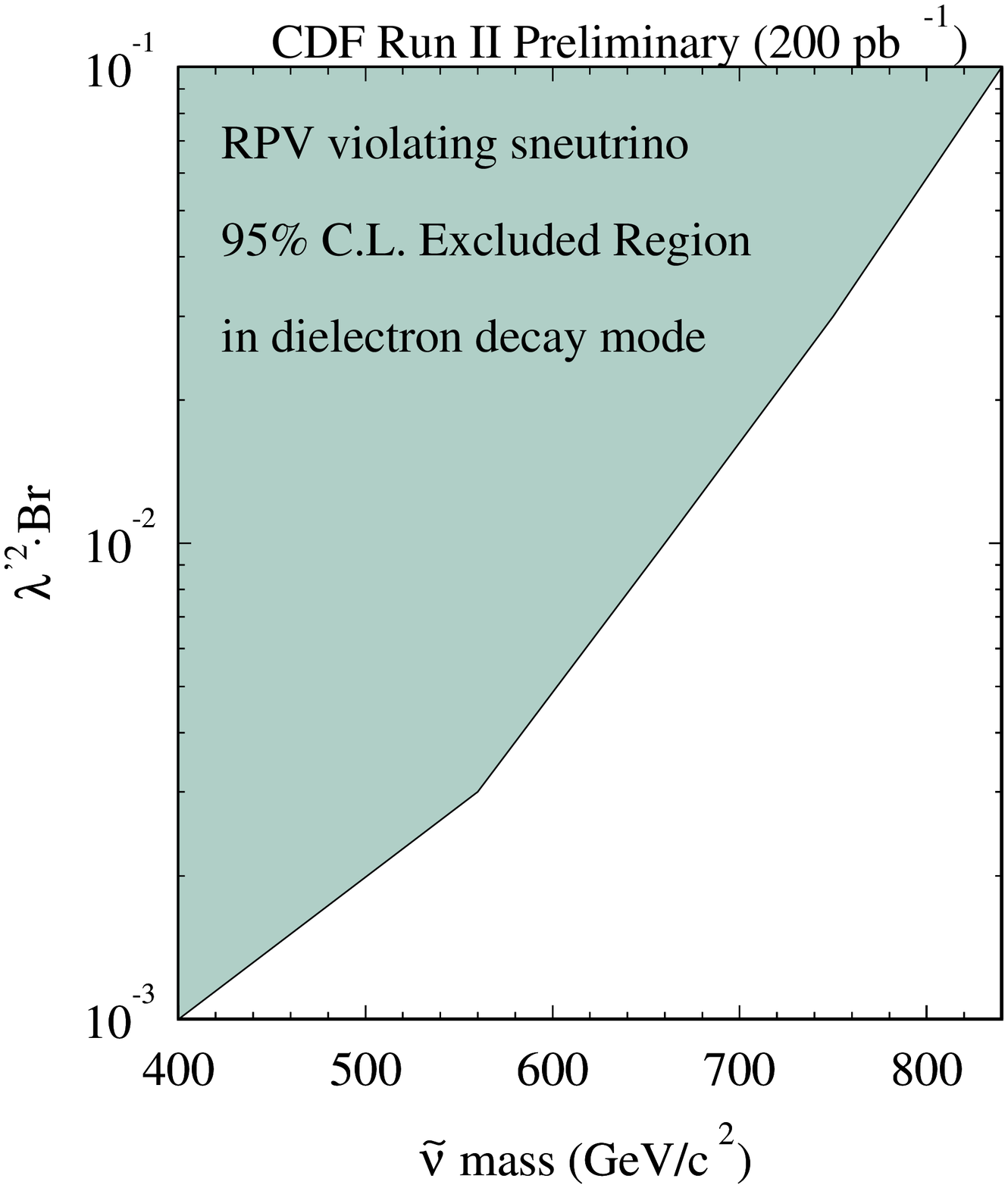}
        }
        }
 \caption{\it
(a) Exclusion limits on $\lambda'_{1j1},\ j=1,2$ as a function of the $s$quark mass.
The two full curves indicate the strongest and the weakest limits. Indirect limits
from neutrinoless double beta decay ($\beta\beta 0\nu$) and atomic parity violation (APV)
are also shown.  
(b) Exclusion of the RPV violating $\lambda'$ coupling as a function 
of the $\tilde\nu$ mass obtained by CDF. $Br$ is the decay branching ratio of the  $\tilde\nu$
into an oppositly charged electron pair.     
    \label{fg:H1RPVExcl} }
\end{figure}

$R$ parity violation introduces 48 new unknown Yukawa couplings, $\lambda_{ijk}$
in the Lagrangian:
\begin{equation}
L = \lambda_{ijk}L_iL_j\tilde E_k + \lambda'_{ijk}L_iQ_j\tilde D_k 
+ \lambda''_{ijk}U_iD_j\tilde D_k, 
\label{eq:RPV}
\end{equation}
where $i,j,k$ are the generation indices and $L,(E)$, $Q,(D)$ are isodoublet (isosinglet)
lepton and quark super-fields, respectively~\cite{SUSY}. At HERA the production
and decay of $s$quarks have been searched for in different final states.
The absence of the expected resonant peak has been transformed to exclusion
of the $\lambda'_{1j1}$ coupling and other SUSY parameters. An example
is shown in Fig.~\ref{fg:H1RPVExcl}$a$~\cite{H1Results}$(ii)$. 
At the Tevatron, CDF searched for a peak in the
opposite sign dilepton distribution as signature of the formation and
decay of $\tilde\nu$~\cite{CDFResults}. The absence of the peak excludes  $\tilde\nu$ masses
and $\lambda'$ parameters as shown in Fig.~\ref{fg:H1RPVExcl}$b$. 

\section{Z'}

Recurrences of the SM gauge bosons appear in several extension of the SM.
For example in the breakdown of the $E_6$ gauge group in addition to $Z_{SM}$
two other $Z'$s appear: $Z_{\psi}$ and $Z_{\chi}$ which can be mixed:
$Z'(\theta) = Z_{\psi}\cos\theta + Z_{\chi}\sin\theta$. 
Depending on the mixing angle $\theta$, different $Z'$ particles are
predicted: $Z_{\psi}$, $Z_{\chi}$, $Z_{\eta}$ and $Z_I$.
Both CDF~\cite{CDFResults} and D\O\ ~\cite{D0Results}$(vii)$  
has searched for signals. In Table~\ref{tb:Zpr}
the limits on the $Z'$ masses are displayed, derived from the
oppositly charged electron pairs.   
\begin{table}
\centering
\caption{ \it $Z'$ mass limits in GeV obtained by the CDF and D\O\ collaborations
}
\vskip 0.1 in
\begin{tabular}{|l|c|c|c|c|c|} \hline
  type &  SM coupling & $Z_I$ & $Z_{\chi}$ & $Z_{\psi}$ & $Z_{\eta}$    \\
\hline
\hline
 CDF   &  750         & 570   & 610        & 625        & 650           \\
 D\O\   &  780         & 575   & 640        & 650        & 680           \\
\hline
\end{tabular}
\label{tb:Zpr}
\end{table}

\section{Little Higgs Model}

The model proposes new fermions and bosons to solve the hierarchy problem.
Contrary to SUSY, here the quadratically divergent diagrams are cancelled
by the same type of particle. $Z_H$ is one of the new bosons to cancel 
divergent boson loops. Its coupling is parametrized by $\Theta$. Using
the absence of the resonant structure at high masses in the oppositly charges
di-lepton pairs CDF has derived the following limits on the 
$Z_H$ mass~\cite{CDFResults}:
$M(Z_H)>800$ GeV for $\cot\Theta$ = 1 (electron pairs) and    
$M(Z_H)>755$ GeV for $\cot\Theta$ = 0.9 (muon pairs).   

\section{Leptoquarks}

\begin{figure}[htbp]
  \centerline{\hbox{ \hspace{0.2cm}
\includegraphics[width=7.0cm]{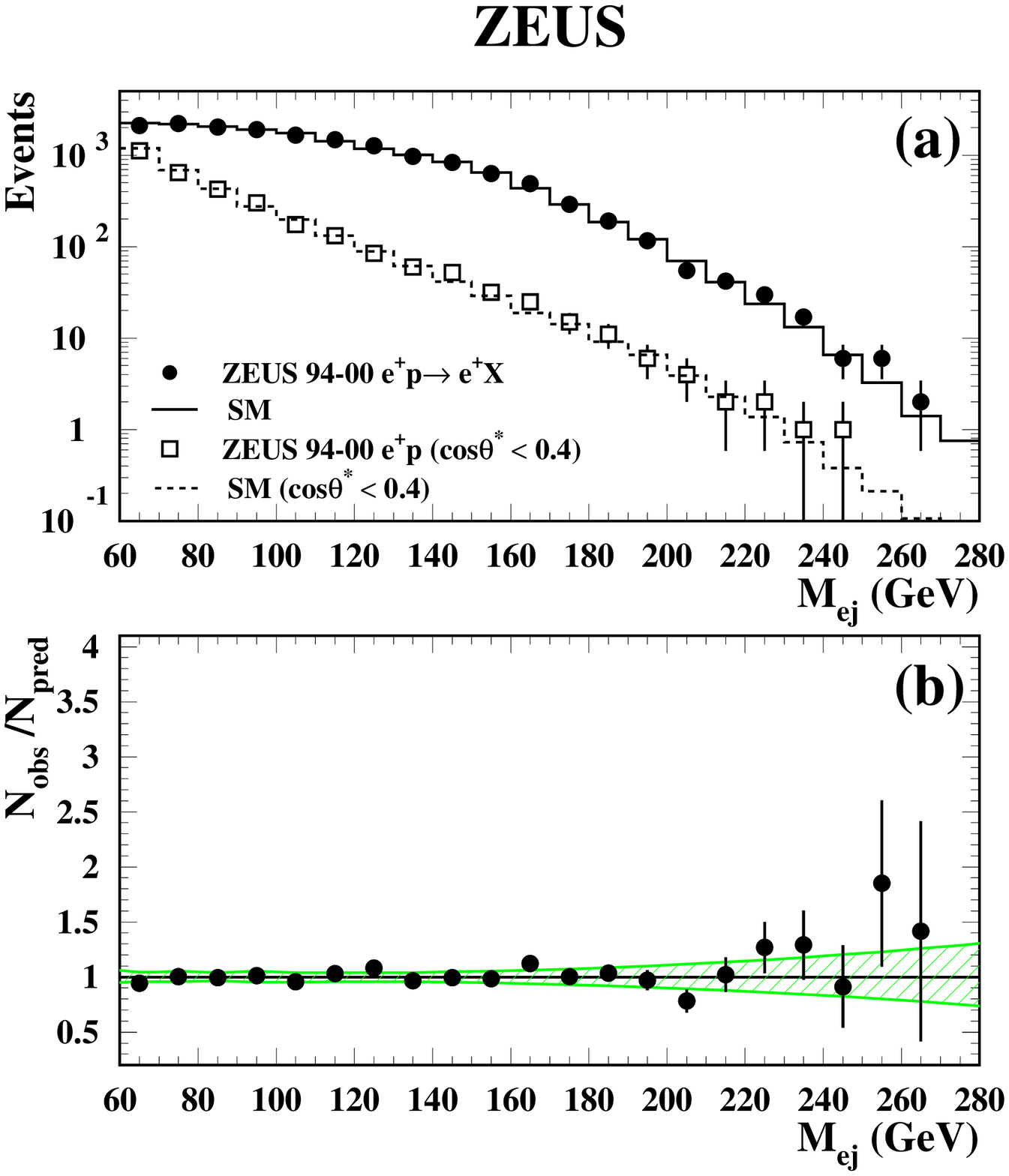}
    \hspace{0.3cm}
    \includegraphics[width=7.3cm]{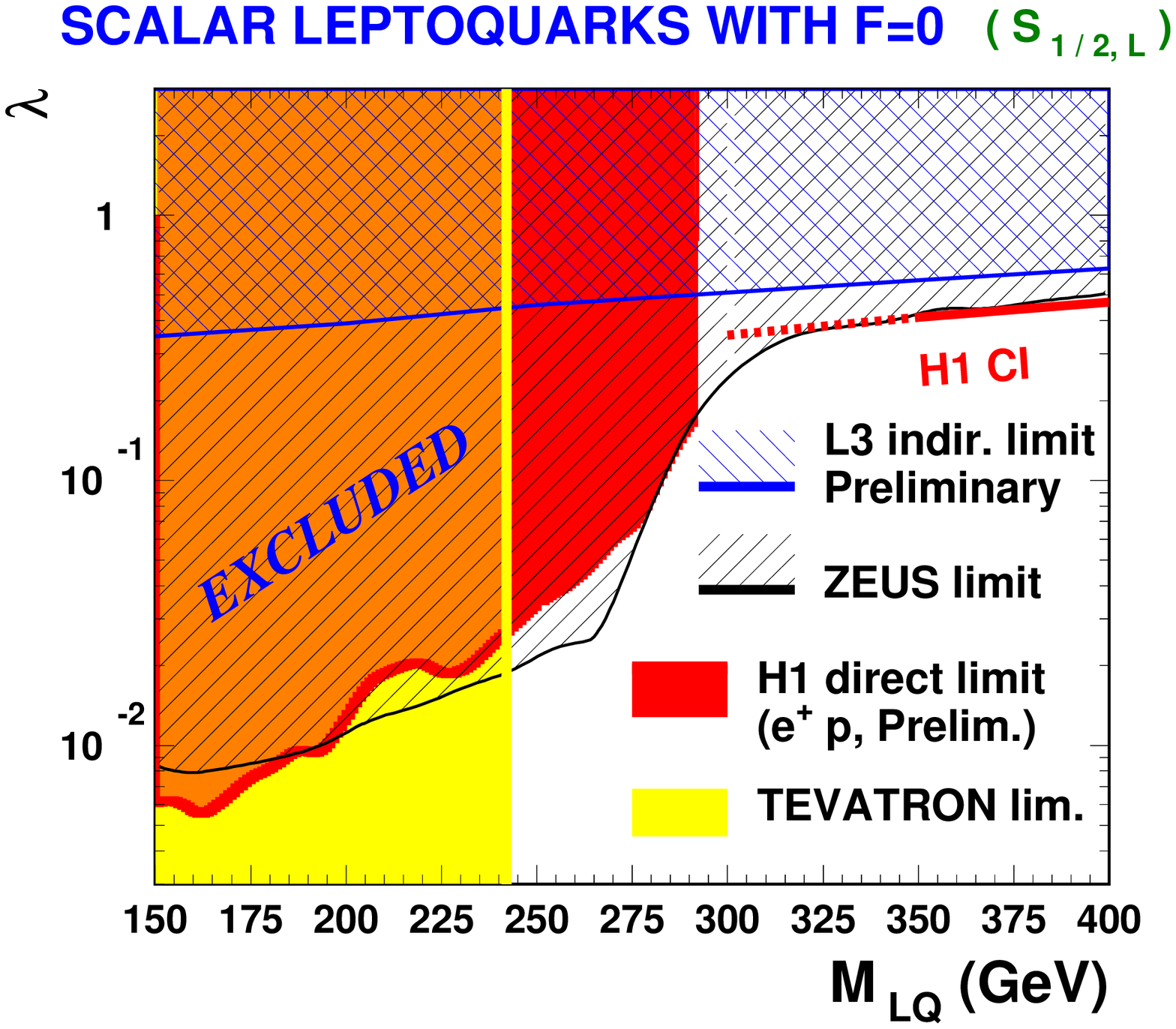}
        }
        }
 \caption{\it
(a) Invariant mass distribution of the final state electron and jet and its
ratio to the SM expectation (left) obtained by the ZEUS collaboration. 
(b) Exclusion region of leptoquarks as function of their mass and coupling.
    \label{fg:ZEUSLQvsSM} }
\end{figure}

Leptoquarks (LQ) are hypothetical bosons (scalars or vectors) which carry both
$L$ and $B$. They are proposed in several extension of the SM based on
quark-lepton symmetry. HERA is an ideal machine to produce 1st generation LQ's,
in the fusion of the incoming electron and quark. The production is proportional
to $\lambda$, the Yukawa coupling of the LQ to the lepton and quark it is composed of.
Leptoquarks, if exist, would show up as resonances in the invariant mass distribution
of the final state lepton and jet. Neither H1~\cite{H1Results}$(iii)$ nor 
ZEUS~\cite{ZEUSResults}$(ii)$ has observed
such signal (see e.g. Fig.~\ref{fg:ZEUSLQvsSM}$a$). This allows
to set limit on the leptoquark mass
depending on its nature (i.e. coupling). Examples of limits are shown
in Fig.~\ref{fg:ZEUSLQvsSM}$b$. 
\begin{figure}[htbp]
  \centerline{\hbox{ \hspace{0.2cm}
\includegraphics[width=6.5cm]{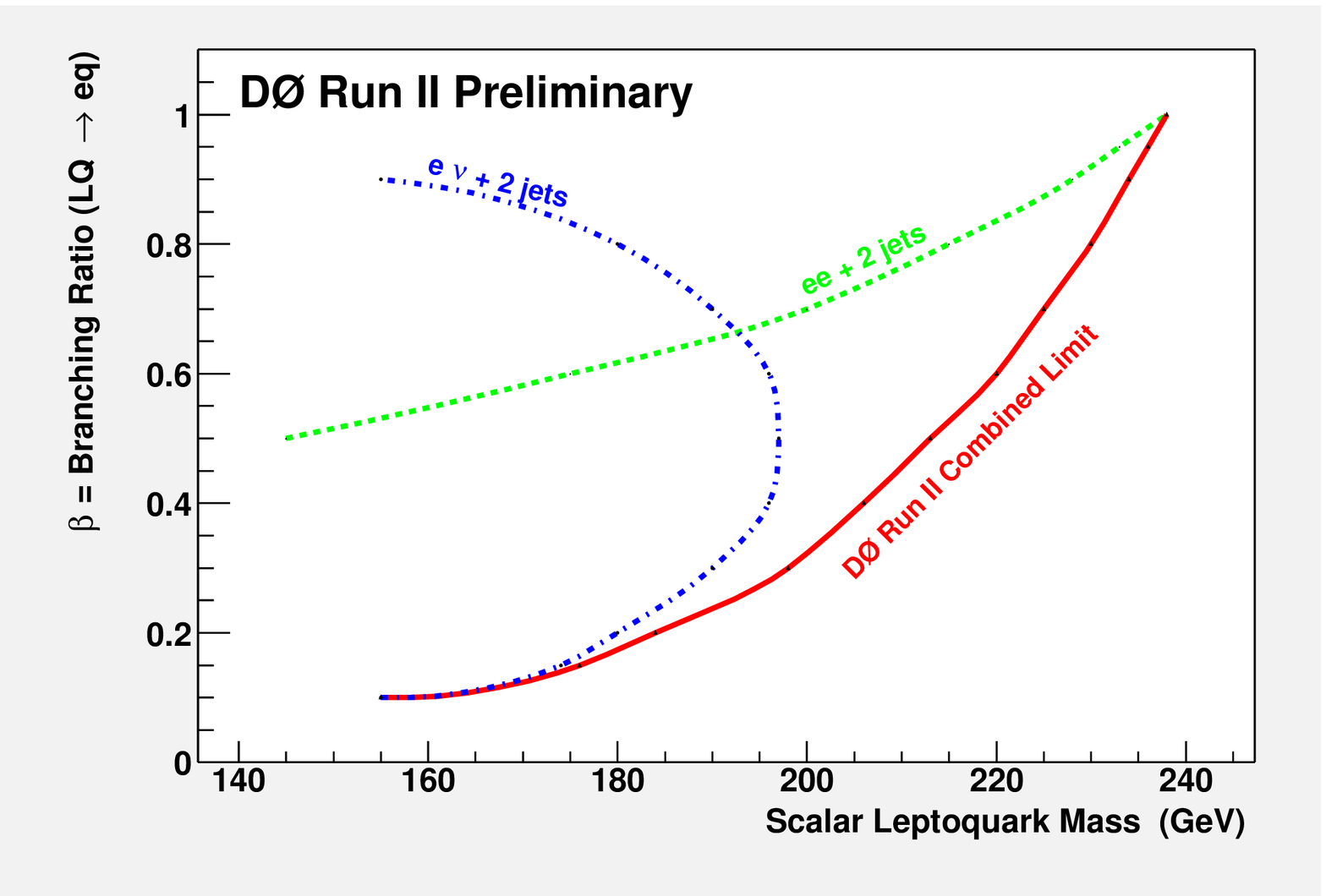}
    \hspace{0.3cm}
    \includegraphics[width=6.8cm]{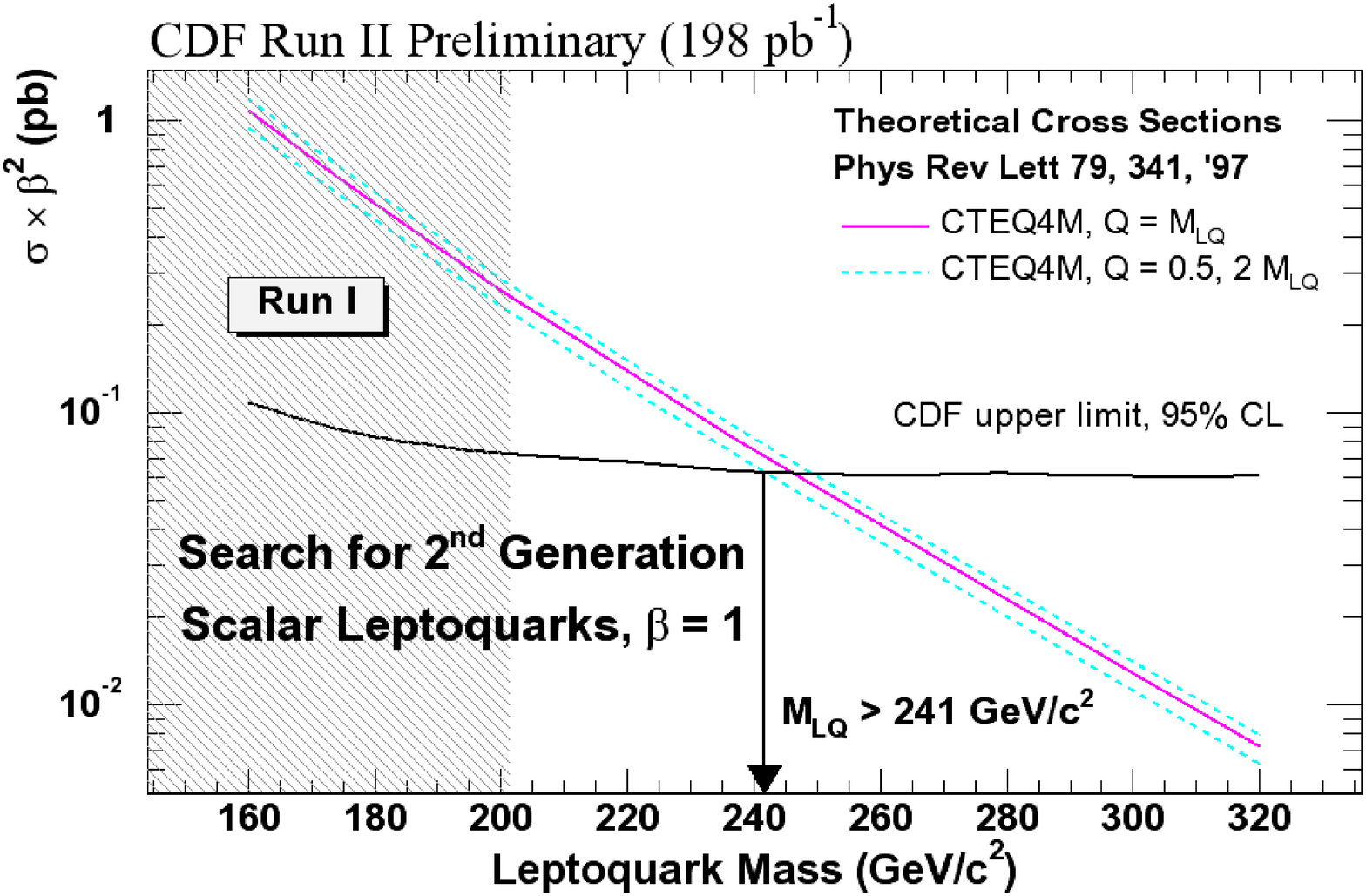}
        }
        }
 \caption{\it
(a) Combined limit on 1st generation leptoquark mass obtained by D\O\ (left). 
(b) Limit on 2nd generation leptoquark mass obtained by CDF (right) with
$\beta=1$.    
    \label{fg:LQ1TevLimitComb} }
\end{figure}

At the Tevatron LQ's are dominantly produced in pairs of the same generation.
The final state is therefore characterized by 2 jets and 2 leptons, where
the leptons can be either charged or neutral, with a branching ratio of $\beta$.
One studies final states with 2 jets + 2 charged leptons, 2 jets + 1 charged
lepton and MET, or 2 jets + MET. The production is independent of $\lambda$. 
The signal dominates the background in regions where the scalar sum of jets
and leptons are large. In this region however the observed number of events is
compatible with that expected from the background. Using that 
both D\O\ ~\cite{D0Results}$(viii)$ and
CDF~\cite{CDFResults} has set new lower limits on the leptoquark mass as a function of $\beta$,
as indicated in Fig.~\ref{fg:LQ1TevLimitComb}$a$ and $b$.

\section{Beyond SM Higgs bosons}

\begin{figure}[htbp]
  \centerline{\hbox{ \hspace{0.2cm}
\includegraphics[width=4.5cm]{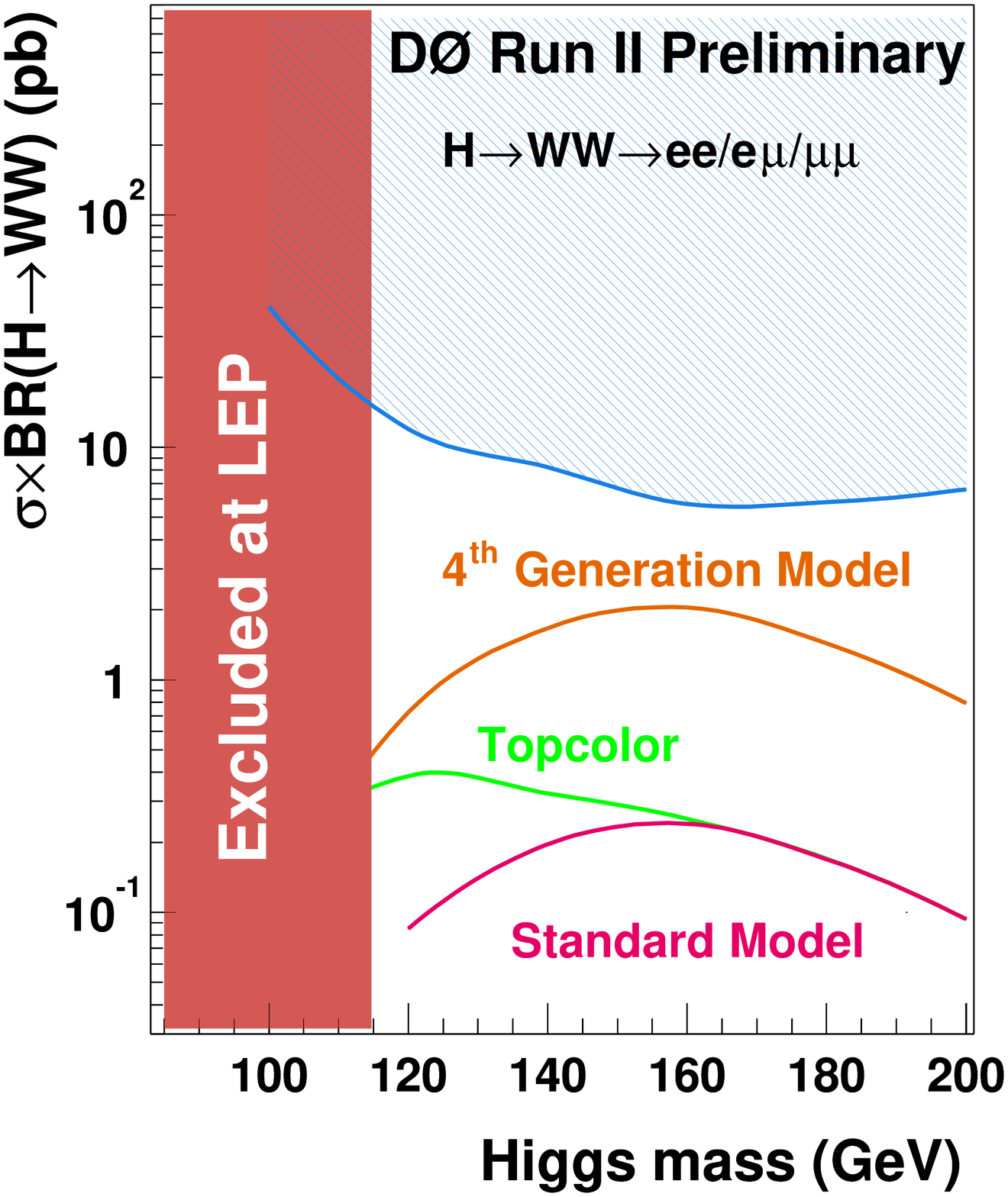}
    \hspace{0.3cm}
    \includegraphics[width=8.0cm]{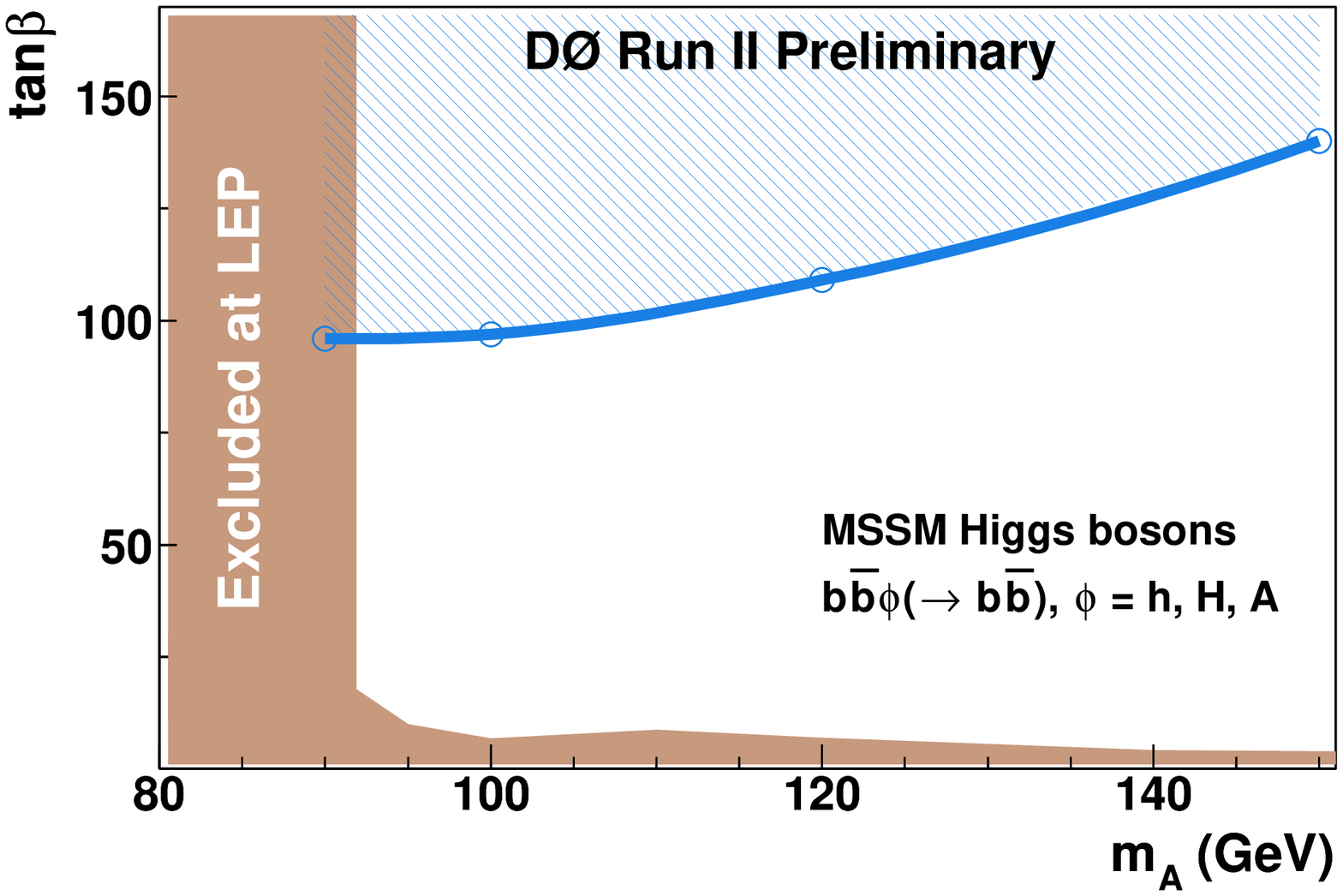}
        }
        }
 \caption{\it
(a) Limit on the Higgs production cross section 
multiplied by the decay branching ratio into $2W$'s,
where both $W$ decays leptonically by D\O\ (left). 
(b) Limit on $\tan\beta$ as a function of the neutral MSSM Higgs mass (right)
obtained by D\O .    
    \label{fg:HWWLimits} }
\end{figure}

Physics beyond the SM may be observed also in the Higgs sector. One searches
either for those Higgses which are not predicted by the SM, e.g. SUSY Higgses,
double charged Higgses, etc, or for SM-type Higgses with anomalous production
cross section or decay rates. D\O\ has established upper limits for branching
ratios for the $h\rightarrow\gamma\gamma$ decay as a function of the
Higgs mass~\cite{D0Results}$(ix)$. CDF~\cite{CDFResults} and 
D\O ~\cite{D0Results}$(x)$ have obtained comparable limits on cross section 
of the Higgs production multiplied by the decay branching ratio into $2W$'s,
where both $W$ decays leptonically (see Fig.~\ref{fg:HWWLimits}$a$). 
D\O\ has set limits on the SUSY parameter $\tan\beta$ as a function of
the mass of the neutral MSSM Higgs boson~\cite{D0Results}$(xi)$ (see Fig.~\ref{fg:HWWLimits}$b$).
Results from searches for multiple lepton final states performed on both colliders
have been interpreted in double charged Higgs bosons scenarios.
Although some excess has been found by H1 in the $2e$
and $3e$ final states compared to the prediction of the SM~\cite{H1Results}$(iv)$ 
(see Fig.~\ref{fg:H1Multielectrons}$a$), only 1 event agreed with the expected topology
of double charged Higgs production and decay. The ZEUS collaboration didn't
observe similar deviation from the SM and neither CDF~\cite{CDFResults} nor 
D\O ~\cite{D0Results}$(xii)$ have
found excess of multilepton events compatible with the double charged Higgs.
All experiments have set lower limit on its mass.  

\section{Substructure of quarks and leptons}

Allowing for internal structure of quarks and leptons goes certainly beyond
the SM. Possible substructure can manifest itself either in excited states of quarks
and leptons, or in their finite size, or in a contact interaction, the
energy scale of which is much larger than the center of mass energy of the reaction.  
\begin{figure}[htbp]
  \centerline{\hbox{ \hspace{0.2cm}
\includegraphics[width=5.7cm]{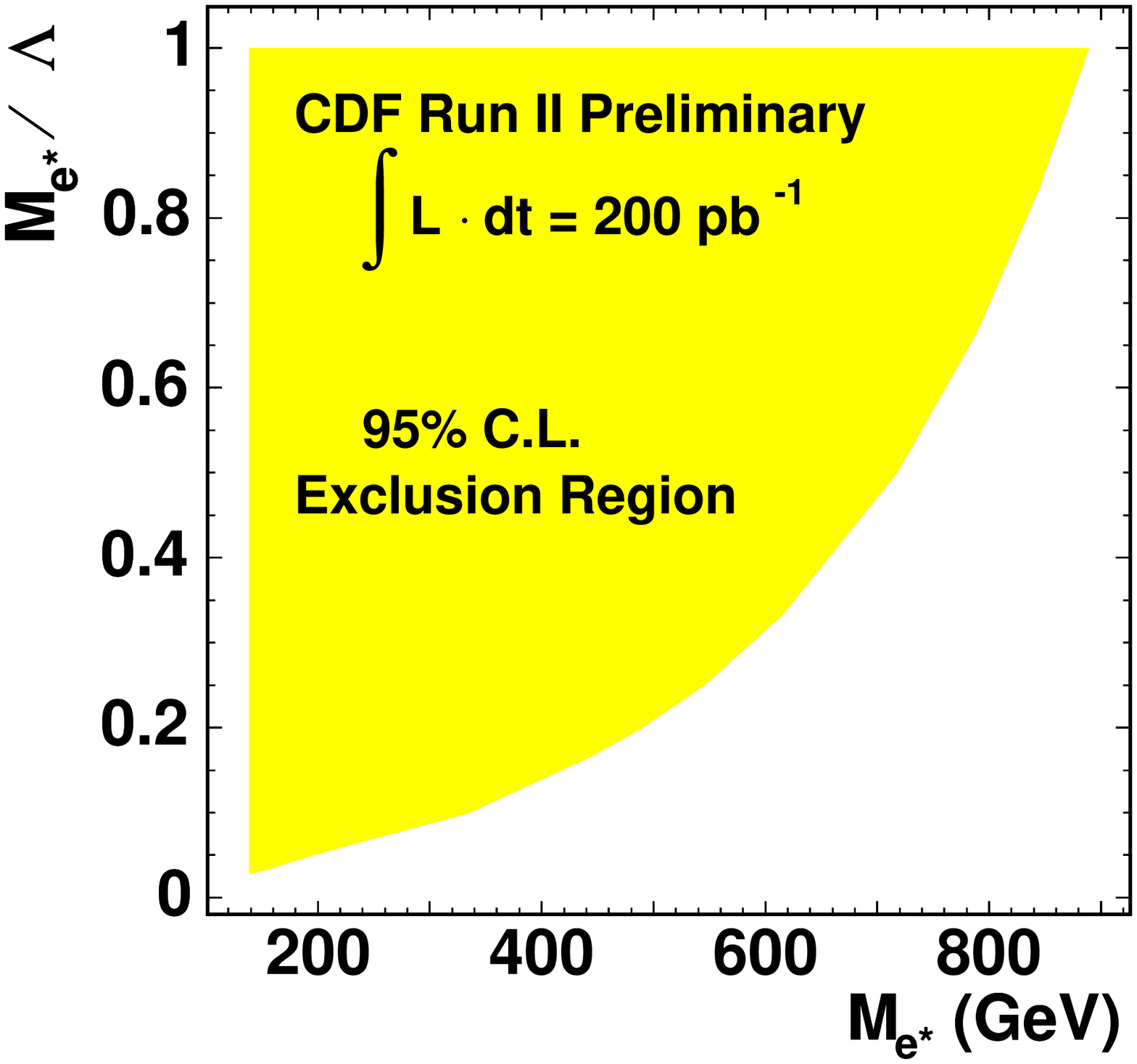}
    \hspace{0.3cm}
    \includegraphics[width=6.2cm]{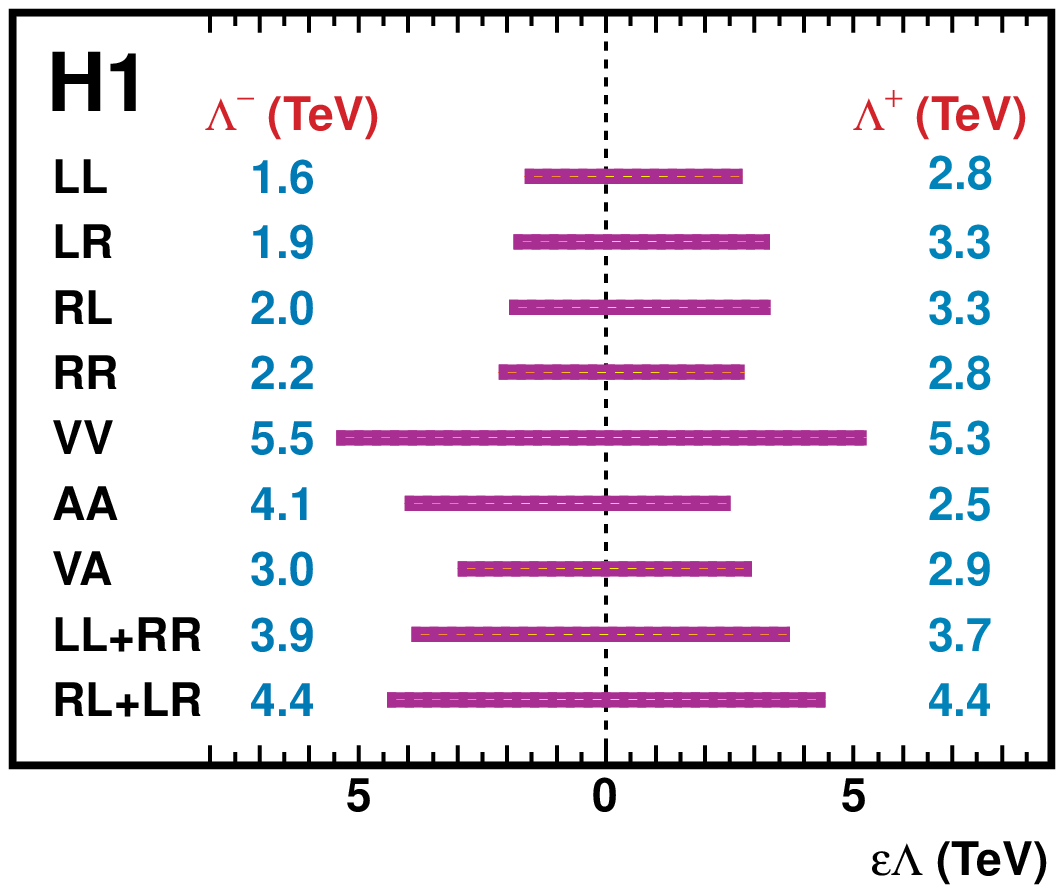}
        }
        }
 \caption{\it
(a) Exclusion region obtained by CDF for the mass of the excited electron, $M_{e^*}$
and the compositness scale, $\Lambda$. (left). 
(b) Lower limits on the compositness scale parameters $\Lambda^{\pm}$ obtained by H1 
in various chiral models (right). $\varepsilon=\pm$ stands for positive or negative
interference with the SM.   
    \label{fg:CDFExcEleLimit} }
\end{figure}
CDF has searched for signal of excited electrons in the final state of two isolated electrons
and an isolated photon~\cite{CDFResults}. 3 events were observed and 3.1 events were expected from the
SM background. This allowed to exclude excited electrons below masses of 889 GeV
depending on the compositness scale $\Lambda$ as shown in Fig.~\ref{fg:CDFExcEleLimit}$a$.

Contact interactions and the size of quarks and leptons have been
searched for at HERA~\cite{H1Results}$(i)$,~\cite{ZEUSResults}$(i)$. 
Both H1 and ZEUS have excluded contact interactions up to the scale
$\Lambda^{\pm}$ of several TeV's, as shown e.g. in Fig.~\ref{fg:CDFExcEleLimit}$b$.
By comparing the observed number of events with that predicted by the SM as a function
of the 4-momentum transfer squared, $Q^2$, both H1 and CDF have set upper limit on the
radius of the quark, $R_q<1.0\cdot 10^{-18}$ m (H1) and $R_q<0.85\cdot 10^{-18}$ m (ZEUS),
assuming that the electron is pointlike.

\section{Searches for anomalies}

\begin{figure}[htbp]
  \centerline{\hbox{ \hspace{0.2cm}
\includegraphics[width=4.5cm]{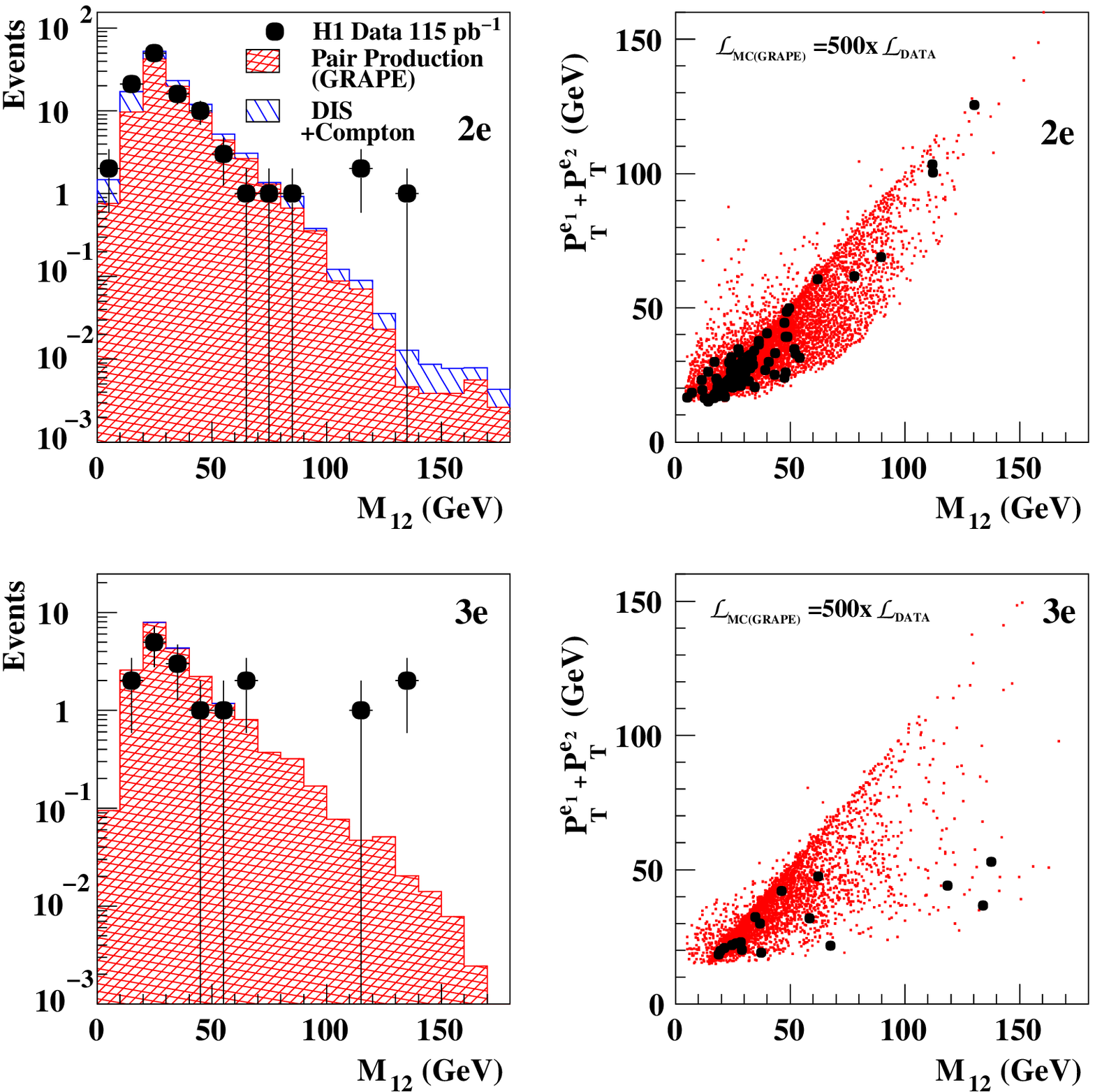}
    \hspace{0.3cm}
    \includegraphics[width=7.0cm]{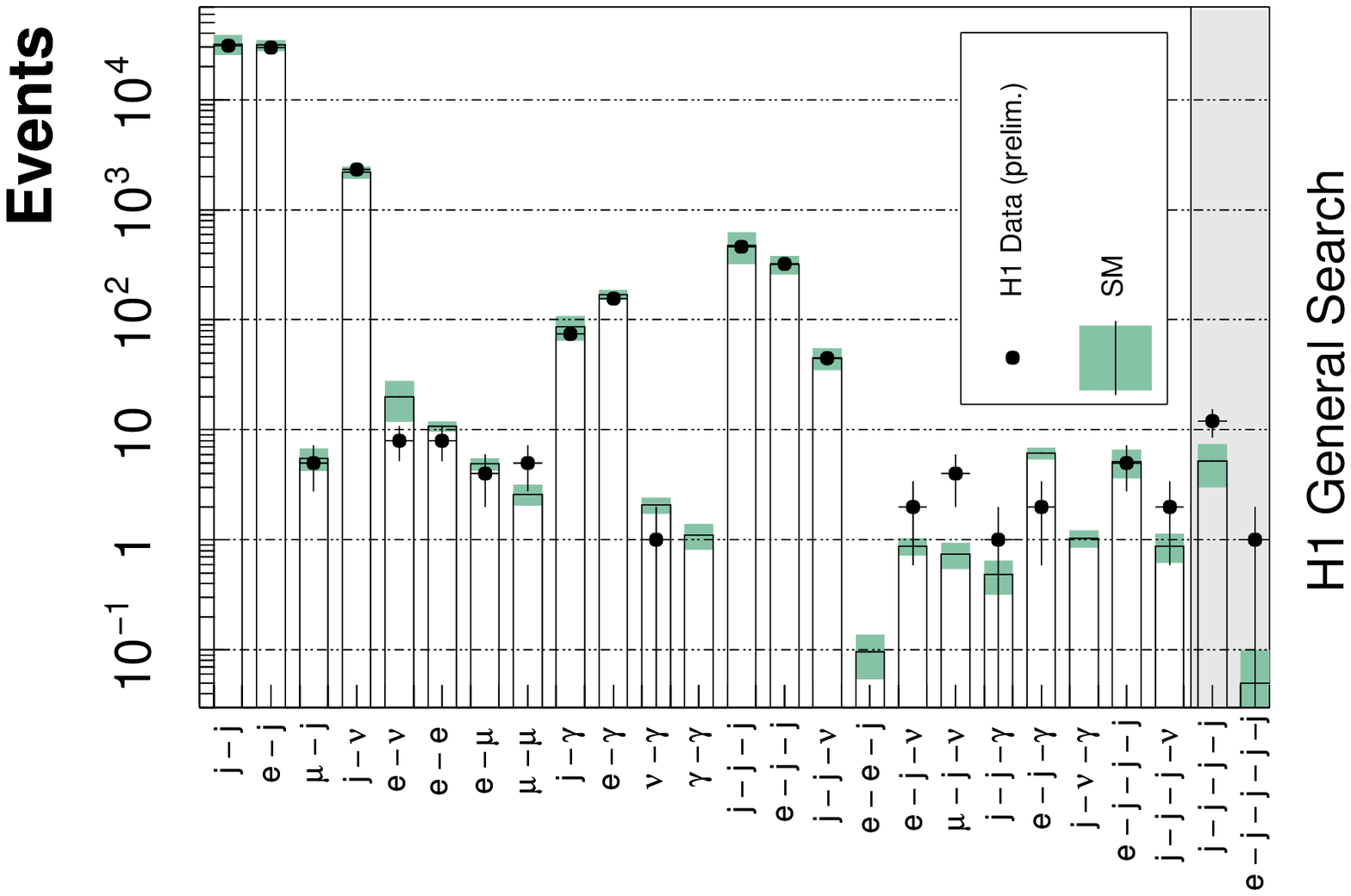}
        }
        }
 \caption{\it
(a)  Distribution of the invariant mass $M_{12}$ of the two highest $p_T$ electrons
(left) and the correlation between $M_{12}$ and the sum of $p_T$'s (middle).
Large size dots are the data, histogram and small size dots represent the
SM (with scaled up luminosity). 
(b)  Number of observed events (data points) and expected number of events
from the SM (squares) in different event categories selected by H1 (right).    
    \label{fg:H1Multielectrons} }
\end{figure}

As already mentioned,  H1 has found 6 multi-electron events of high mass, 
as shown in Fig.~\ref{fg:H1Multielectrons}$a$, in excess to the SM prediction.
At present, there is no explanation for this excess.
In addition, H1 has also found 6 events with an isolated
lepton (3 events with electrons and 3 with muons), accompanied by high MET
and jets of high transverse energy, whereas only 1.3 events are expected
from the SM~\cite{H1Results}$(v)$. 
A possible explanation of these latter events would be single top
production with FCNC coupling. Indeed, 5 events are compatible with this
hypothesis and H1 determines the corresponding inclusive cross section
$\sigma(ep\rightarrow etX)=0.29\pm0.15$ pb. ZEUS does not observe
that phenomenon and it determines exclusion contour for the 
$tu\gamma$ and $tuZ$ FCNC couplings~\cite{ZEUSResults}$(iii)$.  

H1 has also undertaken a systematic, model independent search
for anomalies~\cite{H1Results}$(vi)$. 
It selects events with at least 2 isolated objects among
electrons, muons, photons, jets and MET with $E_T>20$ GeV,
and compares the number of observed events with that predicted
in different bins of their invariant mass distribution or
of their sum of $p_T$ spectrum. As shown in Fig.~\ref{fg:H1Multielectrons}$b$,
significant deviation from the SM has observed only in the $\mu -j-\nu$ channel.

\section{Conclusion}

The performances of both Tevatron and HERA improve steadily
allowing to test new ideas of  ever increasing number 
in the search of the ultimate theory of matter. Although
some anomalies have been observed, no conclusive sign
for new beyond SM physics has appeared yet. More results are expected
soon, already at the late summer conferences of this year.


\section{Acknowledgements}

I am grateful to colleagues of the CDF, D\O , H1 and ZEUS Collaborations
for help in preparing this paper, especially to Elisabetta Gallo
(ZEUS) and Jianming Qian (D\O ). At the same time,
I apologize for subjects which I haven't had space to cover
here.
The highly efficient organization of the conference and the beautiful environment
of Boston University is very much appreciated.


\begin{thebibliography}{99}
\bibitem{D0Results}D\O\  Collaboration 
\url{http://www-d0.fnal.gov/Run2Physics/WWW/results/}\\
$(i)$ D\O\  Note 4336-Conf February 25 2004,
$(ii)$ D\O\  Note 4400-Conf March 19 2004,
$(iii)$ D\O\  Note 4349-Conf March 17 2004,
$(iv)$ D\O\  Note 4368,
$(v)$ D\O\  Note 4380-Conf April 1 2004,
$(vi)$ D\O\  Note 4378-Conf April 13 2004,
$(vii)$ D\O\  Note 4375-Conf March 17 2004,
$(viii)$ D\O\  Note 4401-Conf March 18 2004,
$(ix)$ D\O\  Note 4374-Conf, March 18, 2004,
$(x)$ D\O\  Note 4387-Conf, March 24, 2004,
$(xi)$ D\O\  Note 4366-Conf, March 17, 2004,
$(xii)$ see e.g. poster of M. Zdrazil of this conference.
\bibitem{CDFResults}CDF Collaboration 
\url{http://www-cdf.fnal.gov/physics/exotic/exotic.html}.
\bibitem{H1Results}H1 Collaboration 
\url{http://www-h1.desy.de/} 
$(i)$ DESY 03-052 May 2003,
$(ii)$ DESY 04-025, March 2004,
$(iii)$ H1 0105 for Conf LP03,
$(iv)$ DESY 03-082 July 2003,
$(v)$ DESY 02-224 2002 january 2003, and H1 0079 for Conf LP03, 
$(vi)$ H1 0118 for Conf LP03. 
\bibitem{ZEUSResults}ZEUS Collaboration \\
\url{http://www-zeus.desy.de/physics/exo/ZEUS_PUBLIC/exo_public.html}\\
$(i)$ DESY 03-218 December 2003,
$(ii)$ DESY 03-041,
$(iii)$ Paper 495 for EPS03.
\bibitem{SUSY} For the meaning of the SUSY parameters 
see e.g. ATLAS Detector and Physics Performance 
Technical Design report, Vol II. CERN/LHCC/99-15, ATLAS TDR 15, 25 May 1999.
\end{thebibliography}
\end{document}